\documentclass[preprint2]{aastex631}
\usepackage{amsmath}
\PassOptionsToPackage{breaklinks=true}{hyperref}
\def\chandra    {\emph{Chandra}}
\def\xmm        {\emph{XMM}}

\def\lax{\lesssim}

\received{March 01, 2022}
\revised{June 8, 2022}
\accepted{June 11, 2022}
\submitjournal{ApJ}

\shorttitle{A radio filament in A3562}
\shortauthors{Giacintucci et al.}

\begin{document}
%\special{papersize=8.5in,11in}
%\setlength{\pdfpageheight}{\paperheight}
%\setlength{\pdfpagewidth}{\paperwidth}

\title{A candle in the wind: a radio filament in the core of the A3562 galaxy cluster}

\correspondingauthor{Simona Giacintucci}
\email{simona.giacintucci@nrl.navy.mil}

\author{S. Giacintucci}
\affiliation{Naval Research Laboratory, 
4555 Overlook Avenue SW, Code 7213, 
Washington, DC 20375, USA}

\author{T. Venturi}
\affiliation{INAF - Istituto di Radioastronomia,
via Gobetti 101, I-40129 Bologna, Italy}

\author{M. Markevitch}
\affiliation{NASA/Goddard Space Flight Center,
Greenbelt, MD 20771, USA}

\author{H. Bourdin}
\affiliation{Università di Roma Tor Vergata, Via della Ricerca Scientifica, I00133 Roma, Italy}
\affiliation{INFN, Sezione di Roma 2, Università di Roma Tor Vergata, Via della Ricerca Scientifica, 1, Roma, Italy}

\author{P. Mazzotta}
\affiliation{Università di Roma Tor Vergata, Via della Ricerca Scientifica, I00133 Roma, Italy}
\affiliation{INFN, Sezione di Roma 2, Università di Roma Tor Vergata, Via della Ricerca Scientifica, 1, Roma, Italy}

\author{P. Merluzzi}
\affiliation{INAF – Osservatorio Astronomico di Capodimonte, Salita Moiariello
16, I–80131 Napoli, Italy}

\author{D. Dallacasa}
\affiliation{Dipartimento di Fisica e Astronomia, Università di Bologna, Via
Gobetti 93/2, 40129 Bologna, Italy}
\affiliation{INAF - Istituto di Radioastronomia,
via Gobetti 101, I-40129 Bologna, Italy}

\author{S. Bardelli}
\affiliation{INAF – Osservatorio di Astrofisica e Scienza dello Spazio di
Bologna - via Gobetti 93/3, 40129 Bologna, Italy}

\author{S. P. Sikhosana}
\affiliation{Astrophysics Research Centre, University of KwaZulu-Natal, Durban 4041, South Africa}
\affiliation{School of Mathematics, Statistics, and Computer Science, University of KwaZulu-Natal, Westville 3696, South Africa}

\author{O. Smirnov}
\affiliation{NRAO, PO Box 0, Soccoro, NM 87801, USA}
\affiliation{Department of Physics and Electronics, Rhodes University, PO Box
94, Makhanda 6140, South Africa}

\author{G. Bernardi}
\affiliation{INAF - Istituto di Radioastronomia,
via Gobetti 101, I-40129 Bologna, Italy}
\affiliation{Department of Physics and Electronics, Rhodes University, PO Box
94, Makhanda 6140, South Africa}
\affiliation{South African Radio Astronomy Observatory, 2 Fir Street, Black
River Park, Observatory, Cape Town 7925, South Africa}

\begin{abstract}

Using a MeerKAT observation of the galaxy cluster A3562 (a member of the
Shapley Supercluster), we have discovered a narrow, long and straight, very faint
radio filament, which branches out at a straight angle from the tail of a radio galaxy 
located in projection near the core of the cluster. The radio filament spans 200 kpc 
and aligns with a sloshing cold front seen in the X-rays, staying inside the front in 
projection. The radio spectral index along the
filament appears uniform (within large uncertainties) at $\alpha\simeq -1.5$.
We propose that the radio galaxy is located outside the cold
front, but dips its tail under the front. The tangential wind that blows 
there may stretch the radio plasma from the radio galaxy into a filamentary 
structure. Some reacceleration is needed in this scenario to maintain the radio spectrum uniform.
Alternatively, the cosmic ray electrons from that spot in the tail can
spread along the cluster magnetic field lines, straightened by that same tangential
flow, via anomalously fast diffusion. Our radio filament can provide constraints
on this process. We also uncover a compact radio source at the Brightest
Cluster Galaxy (BCG) that is 2--3 orders of magnitude less luminous than those in
typical cluster central galaxies --- probably an example of a BCG starved of accretion fuel
by gas sloshing.

\end{abstract}

\keywords{galaxies: clusters: general --- galaxies: clusters: individual
 (A3562) --- galaxies: clusters: intracluster medium --- radio continuum: general 
 --- X--rays:galaxies: clusters}

\section{Introduction}\label{sec:intro}

The current generation of sensitive, high-resolution radio interferometers,
such as LOFAR\footnote{LOw Frequency ARray \citep{2013A&A...556A...2V}},
uGMRT\footnote{Upgraded Giant Metrewave Radio Telescope \citep{2017CSci..113..707G}}, 
JVLA\footnote{Jansky Very Large Array \citep{2011ApJ...739L...1P}},
and the Square Kilometer Array (SKA) precursors ASKAP\footnote{Australia SKA
Pathfinder \citep{2021PASA...38....9H}} and 
MeerKAT\footnote{\citep{2016mks..confE...1J,2018ApJ...856..180C}} 
have started revealing new
phenomena that have been hiding under the sensitivity and resolution limits
of earlier instruments. Of particular interest are the strikingly long and
narrow, very faint synchrotron filaments that MeerKAT observations have
discovered in unexpected places in such different objects as the Galactic
Center \citep{2019Natur.573..235H,2022ApJ...925..165H}, between the lobes 
of the radio galaxy ESO 137--006 in the Norma galaxy cluster \citep{2020A&A...636L...1R}, 
emerging from the radio jets of the radio galaxy IC4296 in A3565 \citep{2021ApJ...917...18C}, 
departing from the northern lobe of the radio galaxy 3C40B, and between this latter 
and the nearby radio galaxy 3C40A in A194 \citep{2022A&A...657A..56K}.
%\citep{2021arXiv211105673K}.

Radio filamentary structures have been seen before, for instance, those
embedded within the lobes and tails of radio galaxies \citep[see][for recent examples]{2019MNRAS.488.3416H,2021ApJ...911...56G,2020A&A...634A...9M,2021NatAs...5.1261B}
and in cluster radio relics 
\citep[e.g.,][]{2001AJ....122.1172S,2014ApJ...794...24O,2018ApJ...865...24D,2020A&A...636A..30R,2022A&A...657A..56K}.
MeerKAT now reveals low-surface brightness radio filaments extending outside the radio 
lobes and tails. In fact, the most prominent radio filament in ESO 137--006 is a 
80 kpc-long, collimated and straight thread that connects the two opposing radio lobes, 
bypassing the central nucleus \citep{2020A&A...636L...1R}.

Radio filamentary features extending from, or in the vicinity of, a radio
galaxy have also been found with the VLA 
\citep[e.g., in 3C129, A2256 and in the Perseus cluster;][]{2002AJ....123.2985L,2014ApJ...794...24O,2020MNRAS.499.5791G,2021ApJ...911...56G}
and with LOFAR and GMRT \citep[e.g.,][]{2016MNRAS.459..277S,2017SciA....3E1634D,2019A&A...627A.176C,2020ApJ...897...93B,2021A&A...649A..37B,2022MNRAS.509.1837P,2022arXiv220104591B} at low radio frequencies, 
which are sensitive to the radio emission arising from old cosmic ray electrons.

The nature of these synchrotron threads is unclear. For the MeerKAT radio 
filament between the ESO 137--006 lobes, \cite{2020A&A...636L...1R} suggested
``some sort of reconnection'' in the cluster magnetic field 
%of theintra-cluster medium (ICM) 
around the two radio lobes, possibly caused by its interaction with the lobes themselves. 
While the
polarization properties of the ESO 137--006 threads are still unknown, the radio
filaments in IC4296 \citep{2021ApJ...917...18C} were found to be 
highly polarized with an ordered longitudinal configuration of the magnetic field vectors, 
suggesting coherent magnetic field structures along their entire length ($\sim 50$
kpc). \cite{2021ApJ...917...18C} proposed that these faint synchrotron threads
are created by relativistic electrons escaping from helical Kelvin-Helmholtz
instabilities in the radio jets and suggested that their structure traces
the magnetic field lines that are frozen into the ambient gas of the host
galaxy. \cite{2022A&A...657A..56K} speculated that the radio filaments linked 
to the radio galaxies in A194 may illuminate magnetic 
flux tubes in the ICM 
\citep[e.g.,][]{2012MNRAS.422..704P,2015IAUGA..2258369B,2018SSRv..214..122D}.
The synchrotron emission is proportional to $N_eB^2$, where $N_e$ is the
density of the relativistic electrons and $B$ is the magnetic field
strength, therefore it is natural to expect that such radio filaments map the
structure of the underlying magnetic fields. However, a peculiar
distribution of $N_e$ may also result in filamentary structures. Numerical
simulations have shown, in fact, that ICM ``weather'' flows can advect
the cosmic-ray electrons from Active Galactic Nucleus (AGN) radio lobes and
create filamentary patterns \citep{2021ApJ...914...73Z,2021A&A...653A..23V}.
The structures seen by LOFAR at low radio frequencies may indeed trace the
spatial distribution of aged cosmic rays and the various acceleration
mechanisms in the ICM \citep{2014IJMPD..2330007B,2019SSRv..215...16V}.
These low-brightness radio phenomena are opening a 
new window
onto the physical processes occurring in the hot magnetized ICM, which is
the final repository for the cosmic rays ever produced by AGNs and
cosmological shocks.

Another interesting example of this new phenomenon has recently been
discovered by MeerKAT in A3562 \citep[hereafter V22]{2022arXiv220104887V},
in the core of the Shapley Concentration.
MeerKAT revels a faint radio thread
branching out from the tail of the head-tail radio galaxy J1333-3141 and
spanning a large length of $\sim 200$ kpc.
%This radio thread 
%delineates the southern edge of a diffuse radio halo that permeates the
%cluster central region.} 
As we will show below, the thread appears to be connected to the structure of the cluster
thermal gas. We believe that this specimen can illuminate the nature and
origin of this new phenomenon and potentially provide constraints on the ICM
physics.

In this paper, we present an imaging and spectral analysis of this radio 
filament using the new MeerKAT data, and complement it with a re-analysis of the
archival X-ray {\em Chandra}\/ and {\em XMM-Newton}\/ observations.  We also
study the faint radio emission associated with the Brightest Cluster Galaxy
(BCG) and its connection to the properties of the surrounding thermal gas.

We use a $\Lambda$CDM cosmology with H$_0$=70 km s$^{-1}$ Mpc$^{-1}$,
$\Omega_m=0.3$ and $\Omega_{\Lambda}=0.7$.  At the redshift of A3562
($z=0.049$) $1^{\prime\prime}$ corresponds to $\sim 1$ kpc.
All errors are quoted at the $68\%$ confidence level. The radio spectral 
index $\alpha$ is defined according to $S_{\nu} \propto \nu^{\alpha}$, 
where $S_{\nu}$ is the flux density at the frequency $\nu$.

\section{MeerKAT observations}\label{sec:obs}

The cluster A3562 (RA$_{\rm J2000}$ = 13h 33m 47s, Dec$_{\rm J2000}$ = $-31^{\circ}$ 40$^{\prime}$ 37$^{\prime\prime}$) 
was observed by MeerKAT on 07 July 2019 (AO1, proposal ID SCI-20190418-OS-01) 
for a total of 10 hours on source, with 64 antennas, a total observing bandwidth of 856 MHz, 
and a central frequency of 1284 MHz.
In V22, this observation was complemented with observations centered on the nearby 
SC1329--313 group and on the cluster A3558, taken as part of the MeerKAT Galaxy Cluster 
Legacy Survey \citep[GCLS][]{2022A&A...657A..56K} and MeerKAT scientific commissioning,  
to obtain a mosaic encompassing the whole A3558 complex in the Shapley Concentration core. 
The mosaic 
%has an average noise level of $6$ $\mu$Jy (close to thermal noise)
%for a $7^{\prime\prime}$ beam, and 
is presented in V22 along with an ASKAP 887 MHz image of the same region
and a morphological and spectral study of the extended radio sources in the whole complex.
In this paper, we focus on the region of A3562, the easternmost cluster 
in the A3558-complex chain.

We use the MeerKAT proprietary observations and refer to V22 for a description 
of the data calibration. 
We obtained images at 1284 MHz from the full-band data set (0.9-1.7 GHz)  
with $6^{\prime\prime}$ and $15^{\prime\prime}$ resolution.
%and 6 $\mu$Jy beam$^{-1}$ 
%and 9 $\mu$Jy beam$^{-1}$ noise ($1\sigma$). 
We also obtained pairs of sub-band images, centered at 1070 MHz and 1498 MHz,
at three different resolutions to measure the in-band spectral indices and image 
the spectral index distribution. A summary of these images is given in Table \ref{tab:images}. 
All images 
were produced as described in V22 using {\em wsclean} \citep{2014MNRAS.444..606O, 2017MNRAS.471..301O} 
with a Briggs robust parameter of 0 \citep{1995PhDT.......238B} and fitting a 
fourth-order polynomial to account for wide--band spectral effects. 
No correction for the MeerKAT primary-beam attenuation was applied to the images,
as our analysis is limited to a region within $8^{\prime}$ from the phase center, 
where the effect is negligible  \citep{2020ApJ...888...61M}. The average residual 
amplitude calibration errors are estimated to be of the order of 5\% (V22).

%%%%%%%%%%%%%%%%%%%%%%%%%%%%%%%%%%%%%%%%%%%%%%%%%%%%%%%%%%%%%%%%%%%%%
\begin{deluxetable}{cccc}
\tablecaption{MeerKAT images}
\label{tab:images}
\tablehead{
\colhead{$\nu$} & \colhead{$\Delta\nu$} & \colhead{FWHM}   & \colhead{rms}  \\
\colhead{(MHz)}     &  \colhead{(MHz)}    & \colhead{($^{\prime \prime} \times^{\prime \prime}$)} &  \colhead{($\mu$Jy beam$^{-1}$)}  \\
}
\startdata
1284 & 856 & $6^{\prime\prime}.4 \times 6^{\prime\prime}.0$ & 6  \\
1284 & 856 & $15^{\prime\prime}$ & 9  \\
1070 & 428 & $9^{\prime\prime}.4 \times 8^{\prime\prime}.6$ & 10 \\
1498 & 428 & $7^{\prime\prime}.4 \times 6^{\prime\prime}.7$ & 8  \\
1498 & 428 & $9^{\prime\prime}.4 \times 8^{\prime\prime}.6$ & 8  \\
1070 & 428 & $15^{\prime\prime}$ & 14  \\
1498 & 428 & $15^{\prime\prime}$ & 10   \\
1070 & 428 & $30^{\prime\prime}$ & 40   \\
1498 & 428 & $30^{\prime\prime}$ & 25  \\
\enddata \tablecomments{Column 1: central frequency. Column 2: bandwidth. 
Column 3: Full Width Half Maximum (FWHM) of the radio beam.  Column 4: rms noise level ($1\sigma$).
All images were made using a Briggs robust parameter of 0.}
\end{deluxetable}
%%%%%%%%%%%%%%%%%%%%%%%%%%%%%%%%%%%%%%%%%%%%%%%%%%%%%%%%%%%%%%%%%%%%%

%
\begin{figure*}
\centering
\includegraphics[width=15.5cm]{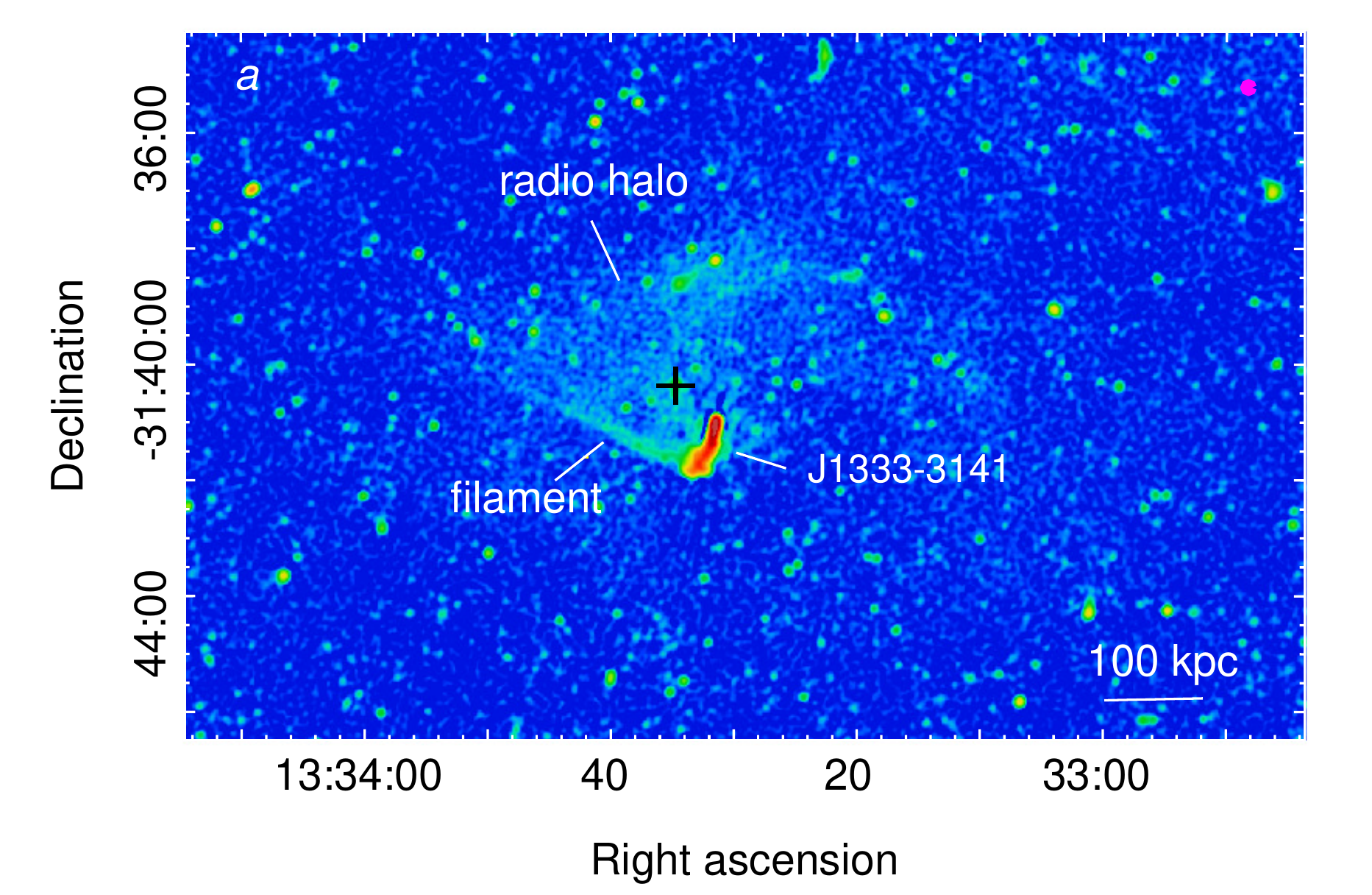}
\includegraphics[width=15.5cm]{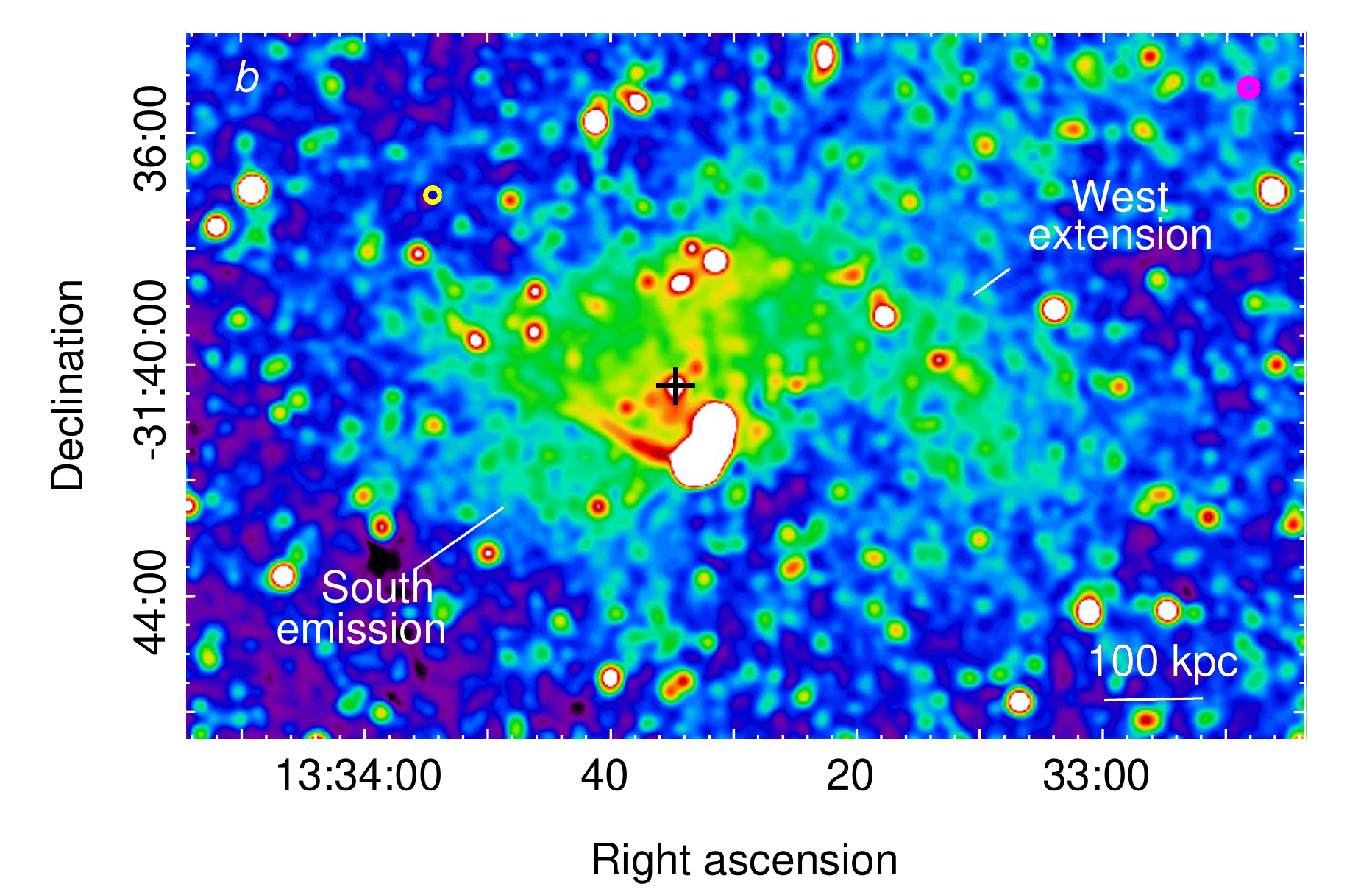}
\smallskip
\caption{MeerKAT 1284 MHz images with an angular resolution of
  $6.^{\prime\prime}4\times6.^{\prime\prime}0$, in p.a. $146^{\circ}$  ({\em a})
  and $15^{\prime\prime}$ ({\em b}). The beam size (FWHM) is shown by a 
  magenta circle in the upper right corner of each panel. The rms noise is
  $1\sigma=6$ $\mu$Jy beam$^{-1}$ and $1\sigma=9$ $\mu$Jy beam$^{-1}$, respectively.
  The black cross marks the position of the BCG.}
  \label{fig:meerkat}
\end{figure*}
%
%%%%%%%%%%%%%%%%%%%%%%%%%%%%%%%%%%%%%%%%%%%%%%%%%%%%%%%%%%%%%%%%%%%%%

%
\begin{figure*}
\centering
\includegraphics[width=13cm]{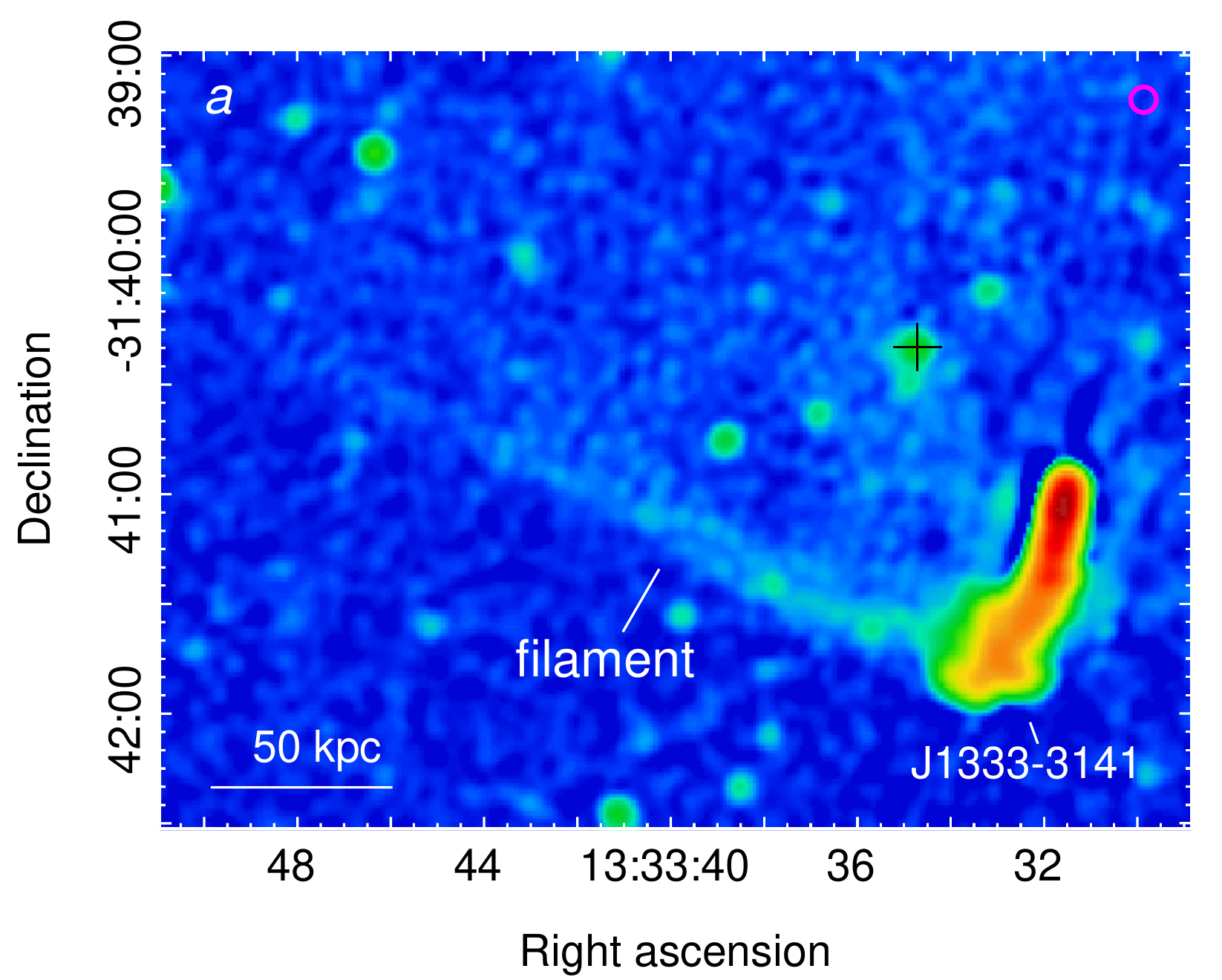}
\includegraphics[width=13cm]{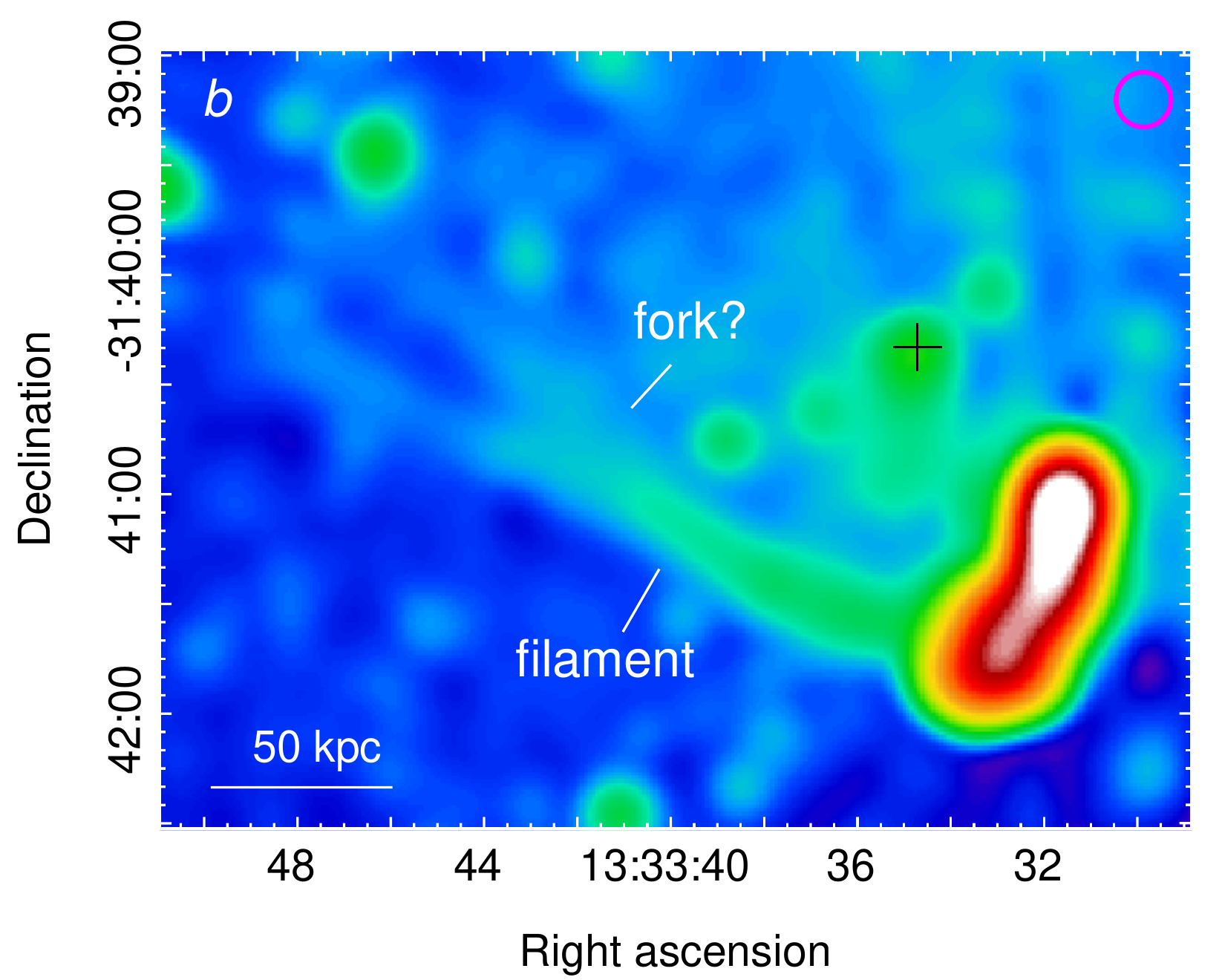}
\smallskip
\caption{MeerKAT 1284 MHz images (same as in Figure \ref{fig:meerkat}) of
the radio filament that starts from the end of the tail of J1333--3141 and
extends $\sim 200$ kpc eastward. A possible fork is seen at $\sim 60$ kpc
from the tail. The beam size is shown by a magenta circle in the upper right corner of
each panel. The black cross marks the position of the BCG.}
  \label{fig:meerkat_fila}
\end{figure*}
%
%%%%%%%%%%%%%%%%%%%%%%%%%%%%%%%%%%%%%%%%%%%%%%%%%%%%%%%%%%%%%%%%%%%%%

\section{MeerKAT view of A3562}\label{sec:images}

In Figure \ref{fig:meerkat}{\em a}, we show the full-band (0.9-1.7 GHz) 
image of A3562 with $6^{\prime\prime}$ resolution from the proprietary data.
In Figure \ref{fig:meerkat}{\em b}, we show
an image of the same region convolved with a $15^{\prime\prime}$ circular beam.
The images have a sensitivity of 6 $\mu$Jy beam$^{-1}$ and 9 $\mu$Jy
beam$^{-1}$, which is almost an order of magnitude increase
in sensitivity over previous best radio observations of A3562 
\citep[hereafter V03 and G05]{2003A&A...402..913V,2005A&A...440..867G}.
%(Venturi et al.
%2003, Giacintucci et al. 2005, hereafter V03 and, G05).

The head-tail radio source J1333--3141 and surrounding diffuse radio halo
were known before \citep[V03, G05,][]{2000MNRAS.314..594V, 2018A&A...620A..25D}. 
The head-tail source is located $\sim
1^{\prime}=60$ kpc (projected) away from BCG 
and is associated with a cluster member galaxy with a radial
velocity $cz=15068$ km s$^{-1}$ that is only $\sim 283$ km s$^{-1}$ 
(about one third of the velocity dispersion, $\sigma=769\pm30$ km
s$^{-1}$) from the cluster mean velocity ($cz=14786$ km s$^{-1}$; Haines
et al. 2018).

The improved sensitivity and angular resolution of the new MeerKAT image
reveals new details and previously-unknown features in A3562, that we
describe in the next sections.

\subsection{The radio filament}

The most striking new feature is the faint, straight radio filament that follows
the southern edge of the bright region of the radio halo in the cluster
core. The radio filament appears to branch out at about a right angle from the
J1333--3141 tail and can be traced for approximately $200$ kpc eastward (see
the zoomed-in images in Fig.~\ref{fig:meerkat_fila}). It is narrow ($\sim 8-10$ kpc) 
and shows a possible fork at $\sim 60$ kpc from the tail, where it seems to split into
two fainter threads. One of them continues roughly along the original
direction, whereas the other seems to depart at an angle of $\sim
15^{\circ}$ from the original axis, merging (in projection) into the halo emission. The radio
filament is also visible in the ASKAP image at 887 MHz
($13^{\prime\prime}$-resolution, V22), but with slightly lower
signal-to-noise ratio due to the sensitivity of the ASKAP observation
($\sim 30$ $\mu$Jy beam$^{-1}$ local noise).

\subsection{The radio halo}

Images of the radio halo at 1.4 GHz (VLA) and at lower frequencies (GMRT)
were presented in V03 and G05, where diffuse emission was detected up to
a largest scale of $500$ kpc along the East-West direction.
In the new, high-sensitivity MeerKAT images (Fig.~\ref{fig:meerkat}),
the halo appears fairly more extended southward (South emission) and
westward (West extension), reaching $\sim 650$ kpc in East-West.

The South emission is a new feature. It is a $300 \times 150$ kpc$^2$ region,
just South of the newly-discovered radio filament, 
with an average surface brightness of $\sim 0.2$ $\mu$Jy/$^{\prime\prime}$
(the average brightness in the halo is $\sim 0.5$ $\mu$Jy/$^{\prime\prime}$).

To the West, the halo shows an extension that can be traced for 
$\sim 300$ kpc. This feature had been partially seen in previous images
(e.g., {\em filament} in Fig.~9 in G05). It appears to point toward the 
neighboring group SC1329--313, believed to have recently interacted with A3562 
\citep[hereafter F04; G05]{2004ApJ...611..811F}. 
Patches of diffuse emission in the region between A3562 and SC1329--313 were
found in previous images (G05), suggesting a possible radio bridge between
the cluster and nearby group. The existence of this bridge was clearly confirmed by
the MeerKAT images presented in V22 (which we also report in Fig.~\ref{fig:arc} in the Appendix),  
where a 800 kpc-long structure connects the 
radio halo to SC1329--313. Along with confirming the radio bridge, 
the V22 MeerKAT mosaic also unveiled a striking, Mpc-scale arc of radio emission 
linking both objects from the North (see also Fig.~\ref{fig:arc}).
Both the bridge and arc have a very low  surface brightness of
$\sim 0.1$ $\mu$Jy/$^{\prime\prime}$ and represent a unique detection of
intra-cluster emission at GHz frequencies, only possible thanks to the exquisite $uv$
coverage and sensitivity of MeerKAT. 
Detections of intra-cluster bridges have been reported at much lower frequencies in the Coma cluster \citep{1989Natur.341..720K,1990AJ.....99.1381V,2011MNRAS.412....2B,2021ApJ...907...32B}
and recently in the cluster pairs A399-A401 \citep{2019Sci...364..981G} and 
A1758N-A1758S \citep{2020MNRAS.499L..11B} with LOFAR at 144 MHz. 

An updated radio spectrum of the halo, combining old (V03, G05)
and new (MeerKAT, ASKAP) flux density measurements has been presented in V22.
Its steep spectral slope
of $\alpha = -1.5\pm0.2$ between 332 MHz and 1284 MHz places the A3562
halo in the class of ultra-steep spectrum radio halos \citep[e.g.,][]{2008Natur.455..944B}.
Because of its relatively compact extent and low radio
luminosity ($P_{\rm 1.4 \, GHz}\sim 1.1\times10^{23}$ W Hz$^{-1}$, V03),
this halo is also one of the smallest and less powerful among the cluster
radio-halo population \citep[e.g.,][]{2019SSRv..215...16V}.

\subsection{The BCG emission}\label{sec:bcg}

\begin{figure}
\centering
\includegraphics[width=10cm]{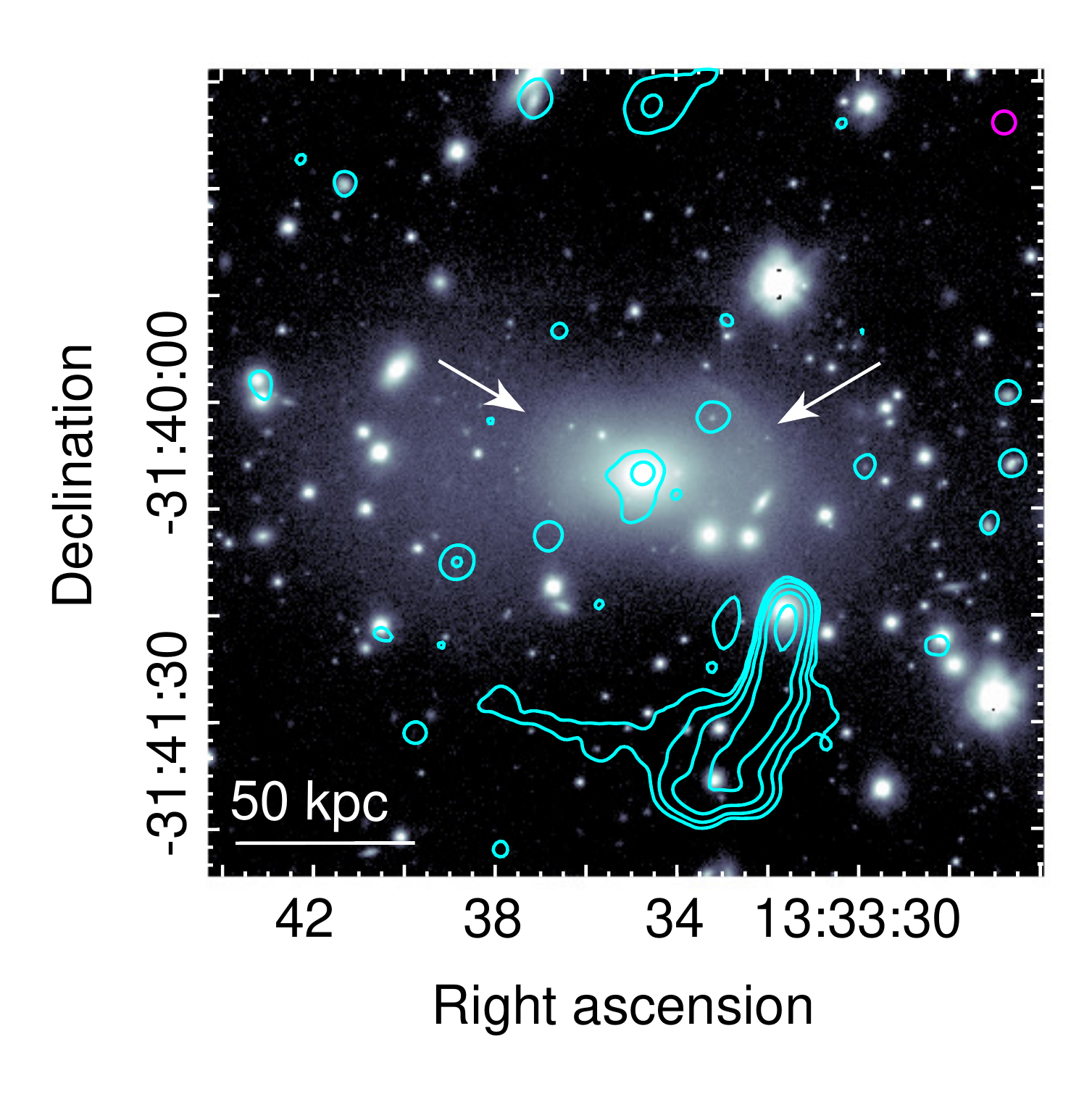}
%\smallskip
\caption{MeerKAT 0.9-1.7 GHz contours (from Figure \ref{fig:meerkat}{\em a})
  overlaid on the VST-OmegaCAM {\em g}--band image of the BCG. Contours start at
  50 $\mu$Jy beam$^{-1}$ and then scale by a factor of 4.
  The radio beam size is shown by the magenta circle.
  Arrows mark shell-like edged (or ripples) in the in the outer regions of the
  galaxy (see also di Gennaro et al.\ 2018).}
  \label{fig:bcg}
\end{figure}

Another feature unveiled by the MeerKAT image of A3562 is weak
radio emission at the position of the BCG. In Fig.~\ref{fig:bcg}, we show
the 1284 MHz radio contours overlaid on the VLT Survey Telescope (VST)-OmegaCAM {\em g}--band
image of the galaxy from the Shapley Supercluster Survey (ShaSS, Merluzzi et al. 2015).
The radio emission consists of a compact component with $0.47\pm0.02$
mJy flux and coincident with the optical peak, and an additional structure 
extending $\sim 10$ kpc southward and with a flux of $0.31\pm0.02$ mJy.
The total radio flux is $0.78\pm0.04$ mJy and the luminosity at 1284 MHz
is $4\times10^{21}$ W Hz$^{-1}$. 

The source was also detected in the ASKAP image at 887 MHz (V22) and by the VLA at 1400 MHz \citep{2005AJ....130.2541M} with unresolved morphology at a resolution of
$13^{\prime\prime}$ and $16^{\prime\prime}$. A very faint compact source
is also visible in the existing GMRT images at lower frequencies \citep[G05,][]{2018A&A...620A..25D}.
No radio source is detected by the Australia Telescope Compact Array at 2360 MHz 
(Venturi et al. 2000), by the VLA at 4860 MHz and 8460 MHz (V03), and in the 
Quick Look images at 3000 MHz from the Epoch 1.2 and 2.1 of the VLA Sky Survey \citep[VLASS;][]{2020PASP..132c5001L}.

In Table \ref{tab:bcg}, we summarize the BCG total flux density and $3\sigma$ upper limits at all frequencies\footnote{a negative artifact affects the BCG location on the GMRT image at 327 MHz, 
therefore we were not able to measure a reliable flux at this frequency.}. 
For MeerKAT, we report the flux density measured on the full-band image at 
1284 MHz (Fig.~\ref{fig:bcg}) and on the two high-resolution 
sub-band images at 1070 MHz and 1498 MHz  (Tab.~\ref{tab:images}). Flux density errors include local noise 
level and residual amplitude uncertainties, as reported in the references.

As visible in Fig.~\ref{fig:bcg}, the radio emission is associated with 
the nucleus of the galaxy and is entirely embedded within the optical halo, which is very
extended ($\sim 180$ kpc) and asymmetrical in East--West. The white
arrows mark the location of sharp,
shell-like edges 
\citep[or {\em ripples}, e.g.,][]{1988ApJ...328...88S,2013pss6.book....1B}
in the outer regions of the galaxy \citep[see also][]{2018A&A...620A..25D}. 
This kind of structures are believed to be late-stage remnants
of a merger between a large elliptical and a smaller companion
disk galaxy \citep[e.g.,][]{1984ApJ...279..596Q}.

%
%%%%%%%%%%%%%%%%%%%%%%%%%%%%%%%%%%%%%%%%%%%%%%%%%%%%%%%%%%%%%%%%%%%%%

%%%%%%%%%%%%%%%% Begin Table BCG %%%%%%%%%%%%%%%%%%%%

\begin{table}[t]
\caption{BCG radio flux densities}
\begin{center}
\begin{tabular}{ccc}
\hline\noalign{\smallskip}
\hline\noalign{\smallskip}
Frequency & Flux density & Reference \\
 (MHz)    &   (mJy)      &          \\
\noalign{\smallskip}
\hline\noalign{\smallskip}
235 & $4.7\pm1.5$ &  G05 \\
610 & $1.4\pm0.3$ &  G05\\
887 &  $1.22\pm0.07$ &  V22\\
1070 & $0.93\pm0.05$ & this work \\
1284 & $0.78\pm0.04$ & this work \\
1498 & $0.62\pm0.03$ & this work \\
1400 & $0.51\pm0.09$ &  \cite{2005AJ....130.2541M}  \\
2360 & $<0.5$        & \cite{2000MNRAS.314..594V} \\
3000 & $<0.4$        &  VLASS \citep{2020PASP..132c5001L} \\
4860 & $<0.2$        & V03 \\
8460 & $<0.1$        & V03 \\
\hline\noalign{\smallskip}
\end{tabular}
\end{center}
\label{tab:bcg}
      {\bf Notes.}
      Flux density errors include local noise level and residual amplitude uncertainties, as
      reported in the references. 
\end{table}

%%%%%%%%%%%%%%%%%%%%%%%%%%%%%%%% End Table BCG %%%%%%%%%%%%%%%%%%%%%%%%%%%%%%%%%%%%%%%%%%%%%%

%%%%%%%%%%%%%%%%%%%%%%%%%%%%%%%%%%%%%%%%%%%%%%%%%%%%%%%%%%%%%%%%%%%%%

%\section{Radio spectral analysis}%

\begin{figure*}
\centering
\includegraphics[width=12cm]{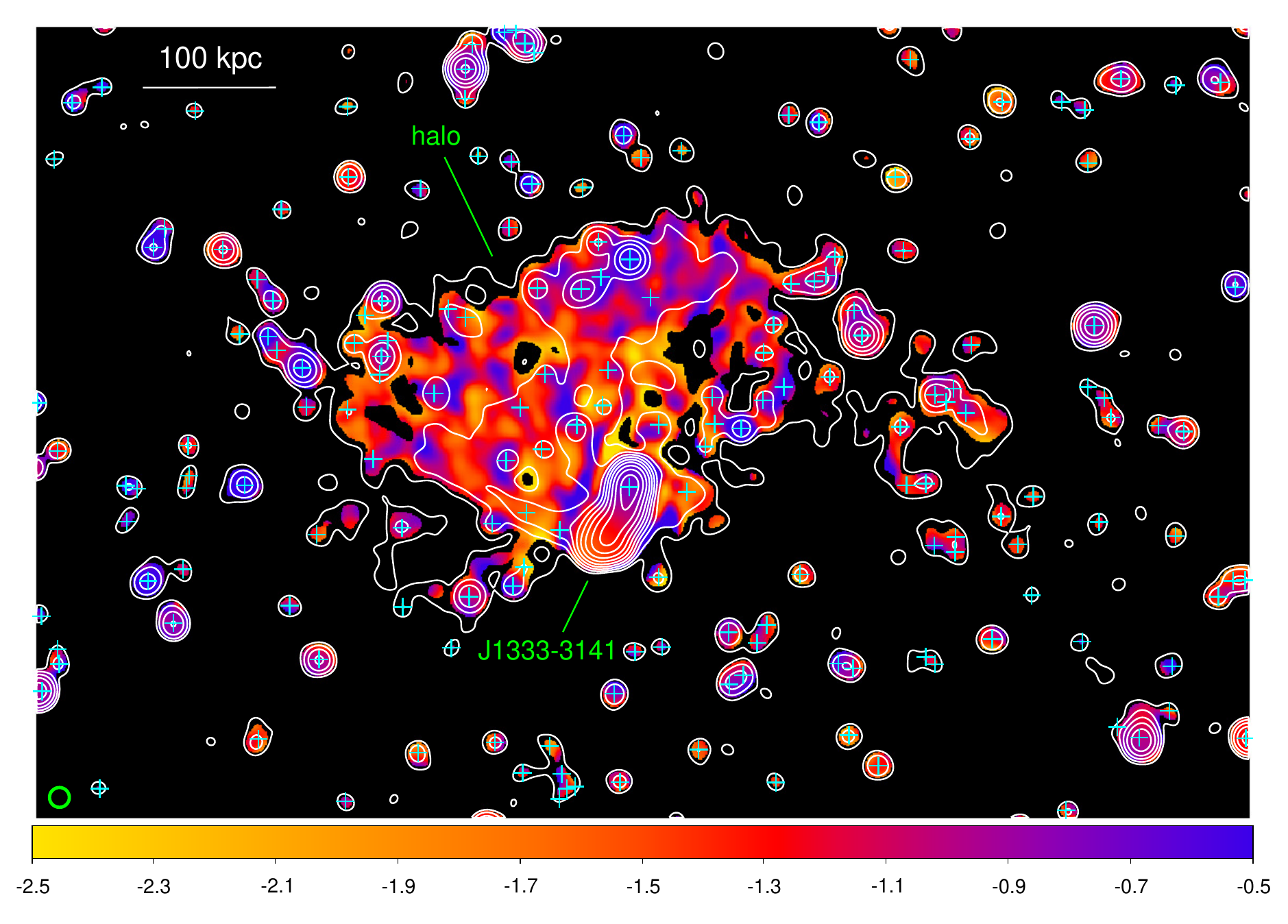}
\includegraphics[width=12cm]{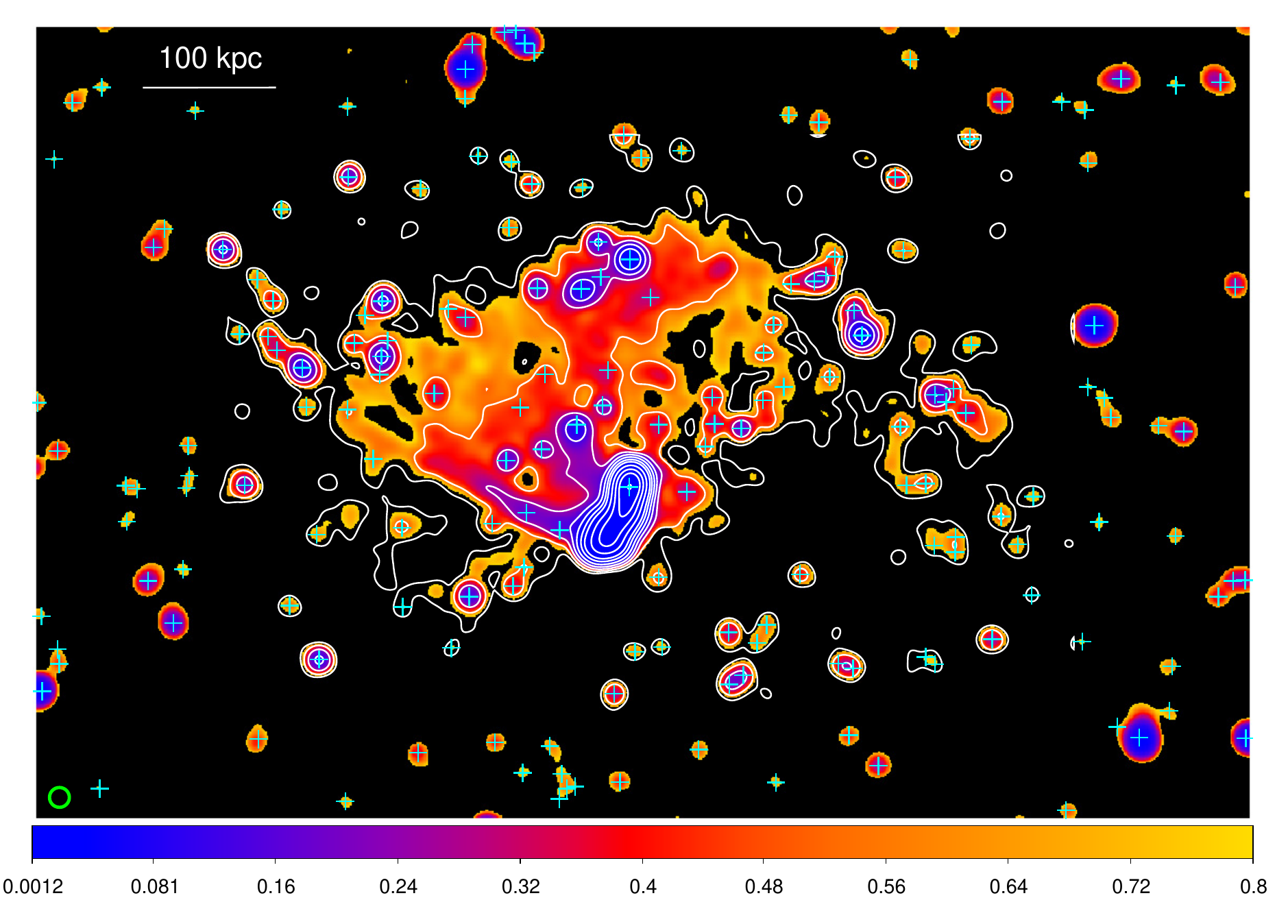}
\smallskip
\caption{Color-scale images of the MeerKAT in--band (1070-1498 MHz)
  spectral index distribution in the head tail J1333--3141 and radio halo
  at a resolution of $15^{\prime\prime}$ (top) and corresponding 
  uncertainty map (bottom). Color bars are the spectral index (top)
  and spectral index error (bottom). In both panels, the contours from 
  the full-band image at matching resolution (Fig.~\ref{fig:meerkat}b) 
  are overlaid on the color map. Contours start at $5\sigma=45$ $\mu$Jy 
  beam$^{-1}$ and scale by a factor of 2. 
  Cyan crosses mark the position of the discrete radio sources identified by the
  PyBDSF source finding on the full-band, high-resolution image 
  (Fig.~\ref{fig:meerkat}a). The beam size is shown by the green circle
  in the left-bottom corner of each panel.}
  \label{fig:spix_high}
\end{figure*}

\begin{figure*}
\centering
\includegraphics[width=12cm]{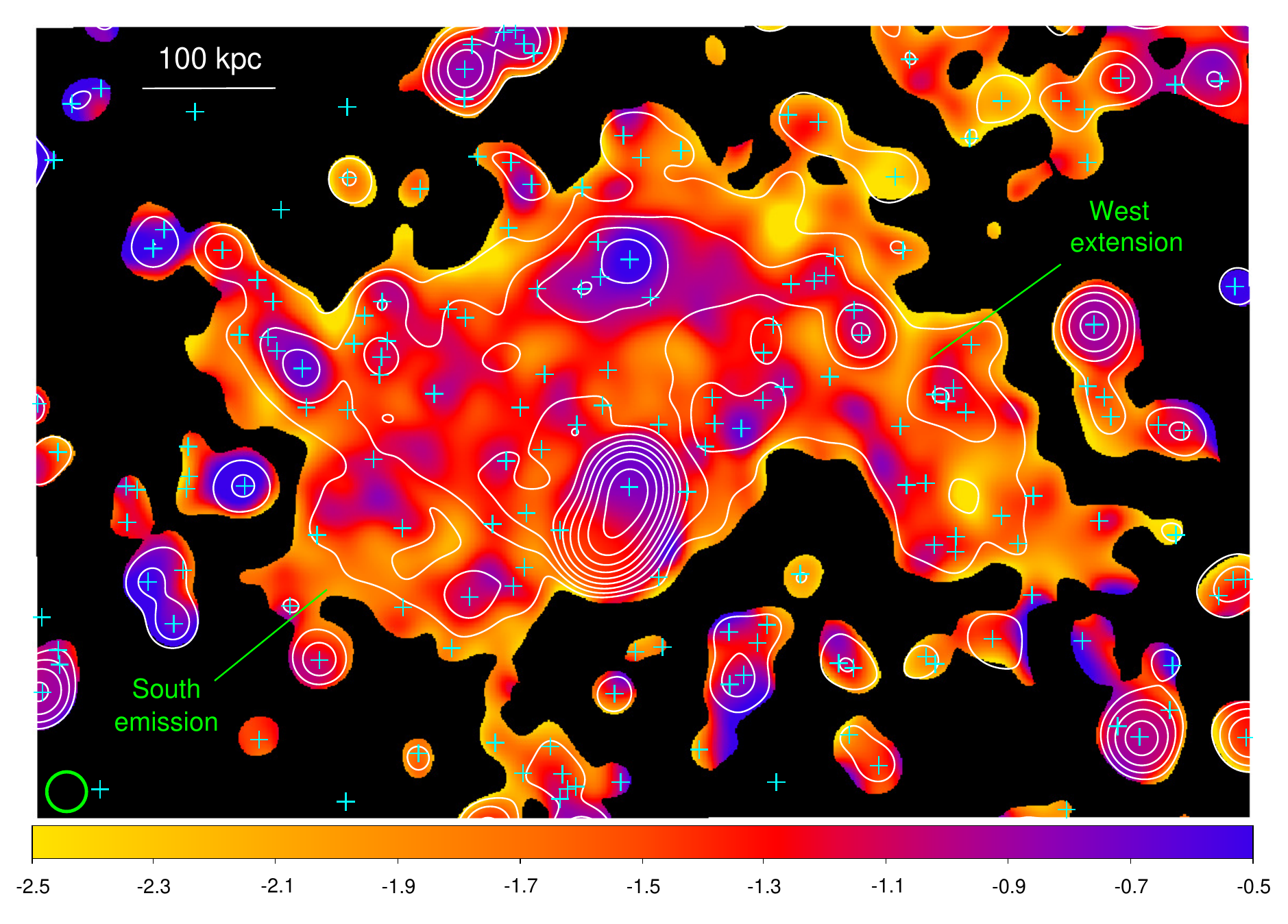}
\includegraphics[width=12cm]{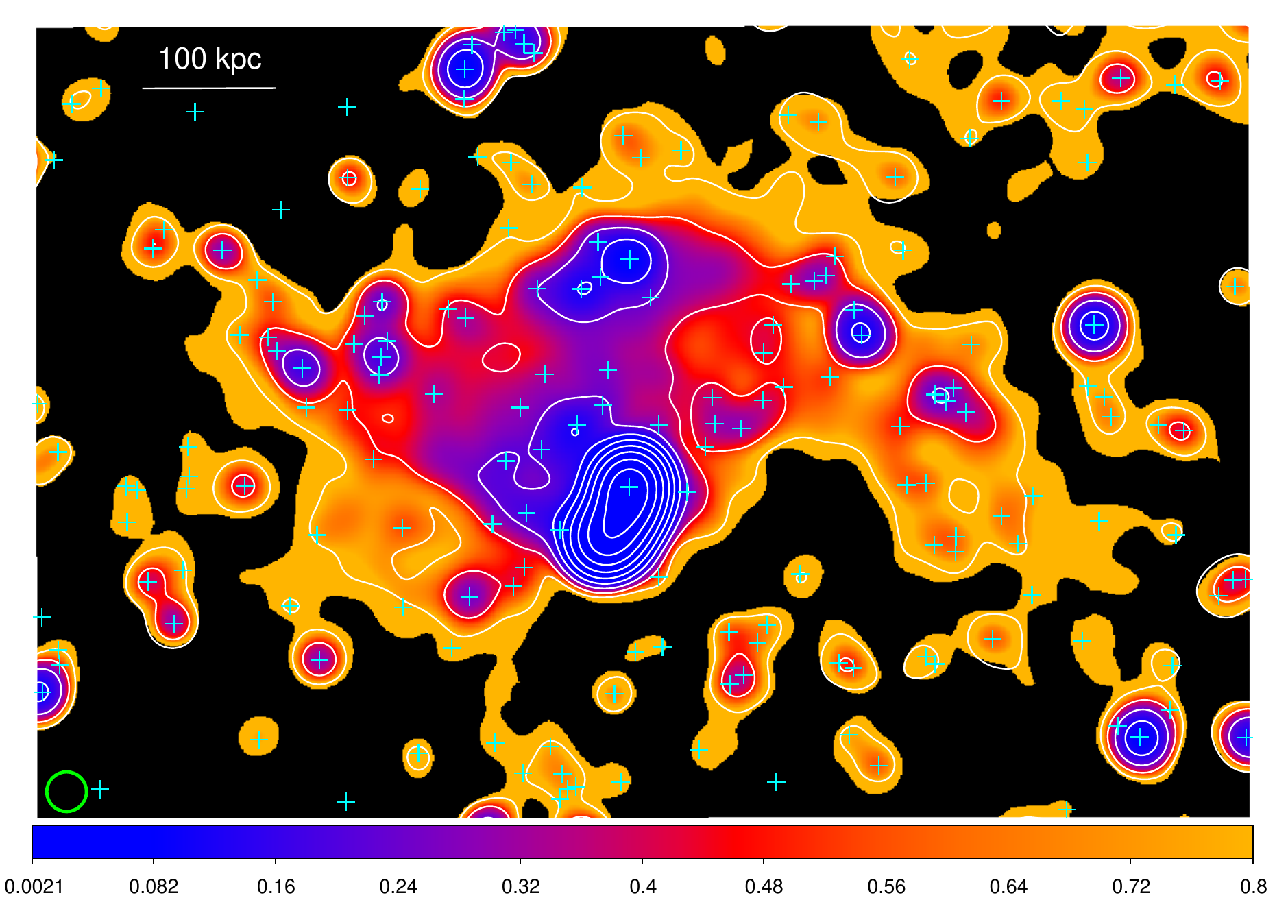}
\smallskip
\caption{Color-scale images of the MeerKAT in--band (1070-1498 MHz)
  spectral index image at a resolution of $30^{\prime\prime}$ (top) and corresponding 
  uncertainty map (bottom). Color bars are the spectral index (top)
  and spectral index error (bottom). In both panels, the contours from 
  the full-band image at matching resolution are overlaid on the color map. 
  Contours start at $5\sigma=75$ $\mu$Jy beam$^{-1}$ and scale by a factor of 2.
  Cyan crosses mark the position of the discrete radio sources identified by the
  PyBDSF source finding on the full-band, high-resolution image (Fig.~\ref{fig:meerkat}a). 
  The beam size is shown by the green circle
  in the left-bottom corner of each panel.}
  \label{fig:spix_low}
\end{figure*}

\section{Radio spectral analysis}

G05 examined the distribution of the spectral index over the radio
halo region using a 332-1400 MHz spectral index map with a relatively
coarse resolution of $\sim 40^{\prime\prime}$. Here we use the 
sensitivity and higher-resolution of the new MeerKAT data to study 
the spectral index distribution in greater spatial detail.

\subsection{Spectral index images}\label{sec:spix}

We computed a MeerKAT in-band spectral index map by comparing a pair of
images centered at 1070 MHz and 1498 MHz, convolved with a $15^{\prime\prime}$ 
circular beam (Tab.~\ref{tab:images}). The spectral
index was calculated in each pixel where both images have a signal above
the $5\sigma$ level. The resulting spectral index map is shown in colors in Fig.~\ref{fig:spix_high}
(top) with $5\sigma$ contours from Fig.~\ref{fig:meerkat}b overlaid in white.
The bottom panel shows the corresponding spectral index uncertainty map,
in which the main contribution comes from the proximity of the frequencies.
Cyan crosses mark the position of discrete radio sources that were identified
using the full-band, high-resolution image (Fig.~\ref{fig:meerkat}a) and
the Python Blob Detection and Source Finder software \citep[PyBDSF;][]{2015ascl.soft02007M}.
Following the same procedure, we obtained a low-resolution ($30^{\prime\prime}$) 
in-band spectral index map using a $3\sigma$ brightness 
cutoff in the individual 1070 and 1498 MHz images (Tab.~\ref{tab:images}). The spectral 
index image is shown in the top panel of Fig.~\ref{fig:spix_low} and associated uncertainty image 
is shown in the bottom panel.  
Superposed are the $3\sigma$ contours from the full-band image at matching resolution.

Along the $80$ kpc-long head tail J1333--3141 in Fig.~\ref{fig:spix_high}, 
we observe a clear gradual steepening from $\alpha \sim -0.6$ (head)
to $-1.5$ (end of the tail; see also Sect.\ref{sec:radioprof} and Fig.~\ref{fig:spix_profile}). 
A similar trend was also seen in the
spectral index maps in G05. Artifacts due to residual image errors around 
J1333--3141 cause the spectral index values to be unreliable in the 
immediate surrounding of this source, in particular in a very steep 
spectrum region (yellow) around the head and in a pair of low-$\alpha$ 
{\em bars} (dark blue) parallel to the tail in Fig.~\ref{fig:spix_high}. 
Unfortunately, one
of these bars affects the location where the radio filament branches out
from the tail of J1333--3141. Beyond this region, the spectral index of the filament
appears relatively constant along its whole $200$ kpc length, with an average
of $-1.5$. This value is similar to the spectral index in the terminal 
region of the head-tail source, from where the radio filament emerges. 
In Section \ref{sec:radioprof}, we examine in detail the spectral 
behavior along J1333--3141 and the filament.

In the halo region, an average value of $\alpha\sim -1.5$ is observed 
across most of the halo, including the West extension and newly-detected 
South emission, in agreement with the G05 map and with the integrated 
spectrum of the halo in V22. The apparent structure in the spectral index maps within the halo  on $\lax 1^{\prime}$ scales (Figs.~\ref{fig:spix_high} and \ref{fig:spix_low}) is not statistically significant. We searched for variations by deriving spectral index values for a grid of $40^{\prime\prime}\times40^{\prime\prime}$ beam-independent regions, as well as for several larger ($\sim 4-8$ arcmin$^2$) regions, masking the head-tail source, 
filament and discrete radio sources. All values were consistent with the average within statistical uncertainties, which are large due to the limited frequency coverage. Upcoming uGMRT observations at lower frequencies, in combination with the MeerKAT data presented here, will provide better sensitivity to spatial variations of the halo spectral index.

\subsection{Spectral index profiles}\label{sec:radioprof}

\begin{figure*}
  \centering
\includegraphics[width=16cm]{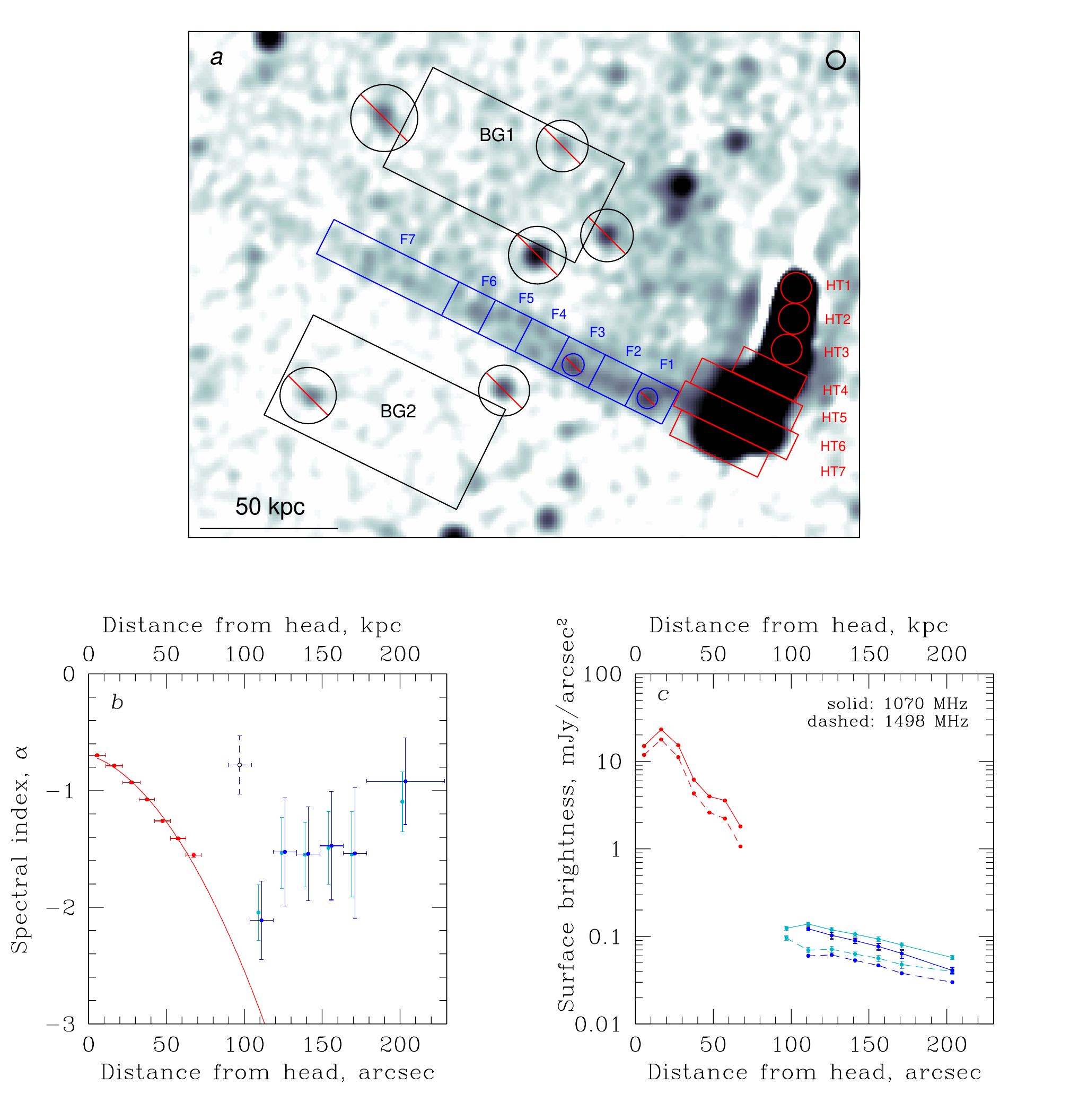}
\smallskip
\caption{MeerKAT 1070-1498 MHz spectral index profiles ({\em b}) 
and surface brightness profiles ({\em c}) at 1070 MHz (solid lines) and
1498 MHz (dashed lines) along the head tail (red) and radio
filament (cyan and blue). The profiles were computed using the regions 
shown in panel {\em a} and overlaid on the full-band image at 1284 MHz. 
The beam size is shown in black in the upper--right corner of the image.
Discrete sources in the F1 to F7, BG1 and BG2 regions were masked out. 
The region F1 (shown in dash in panel {\em b}) is affected by residual image
artifacts near the bright head-tail emission . The filament data points
(except F1) have been corrected for the background halo emission, 
estimated using the BG1 and BG2 regions (see text). Uncorrected data points 
are shown in cyan, corrected points are shown in blue.
The head-tail spectral index profile is well fitted by a JP model (solid
red line), as described in the text.}
\label{fig:spix_profile}
\end{figure*}

We extracted a spectral index profile along the head-tail J1331--3141 and
radio filament using images at 1070 MHz and 1498 MHz with a common beam of
$9^{\prime\prime}.4\times8^{\prime\prime}.6$ (Tab.~\ref{tab:images}). We used
the regions shown in Fig.~\ref{fig:spix_profile}a, i.e., 7 regions along the tail (red) and 8
regions along the filament (blue). The size of each region was chosen to be
larger than at least one beam to sample independent regions.  Since the
filament is projected onto the radio-halo emission that has a spatial 
surface brightness gradient, we used two large regions on two sides 
of the filament (black boxes labelled BG1 and BG2) to estimate the 
background level associated with the diffuse emission from the halo 
and subtract it from the signal in the regions along the filament. 
The image does not have sufficient sensitivity to characterize
the spatial distribution of the underlying emission, so we simply assumed a
linear interpolation between the background regions. 

Figure~\ref{fig:spix_profile}b shows the spectral index profile along the
head-tail source in red. The profile along the filament before the background
correction is shown in cyan and that with background correction is shown 
in blue. Errors were computed taking into account the local 
noise level and uncertainty in the background subtraction.   
The corresponding surface brightness at two frequencies in the same regions 
are shown in Fig.~\ref{fig:spix_profile}c. The emission in the filament has a surface brightness much lower than anywhere in the head-tail source.
The background correction to the filament spectral index becomes noticeable in
the last region of the filament (F7). Selecting different background regions (i.e., closer to the southern edge of the filament; the northern background region could not be moved closer because of the fork, Fig.~\ref{fig:meerkat_fila}b) 
changes the resulting spectral indices by a small fraction of the statistical uncertainties. 

The profile along the tail of J1331--3141 is consistent with the
head-to-tail steepening already visible in the spectral index image in
Fig.~\ref{fig:spix_high}. 
We fitted the observed spectral steepening using a Jaffe-Perola (JP) model 
\citep{1973A&A....26..423J}, in which the timescale for continuous isotropization
of the electrons is assumed to be much shorter than the radiative
timescale. We also assumed that the expansion velocity of the source is
constant and that the break frequency $\nu_{\rm break}$ is proportional to
$d^2$, where $d$ is the distance from the core \citep[e.g.,][]{2007A&A...470..875P}. 
The best fit is shown as a solid red line in
Fig.~\ref{fig:spix_profile}b. The model gives an injection spectral index of
$\alpha_{\rm inj}=-0.72\pm0.01$ and $\nu_{\rm break}=2.08\pm0.07$ GHz, in
very good agreement with the results of a similar spectral analysis carried
out by V03 using images at 2.38 GHz and 8.46 GHz. 
%A radiative age of $\sim
%60$ Myr is inferred at the end of the tail assuming a constant magnetic
%field across the whole source of 5 $\mu$G (V03).

While the steepening along the head tail is consistent with radiative aging
of the relativistic electrons, the spectral index along the radio
filament -- assuming it is physically connected to the tail -- does not
follow the prediction of the JP model. Even though the uncertainties are
large, instead of steepening, the filament spectral index remains relatively
constant around $-1.5$, the value seen in the tail at the position where the
filament branches out (Fig.~\ref{fig:spix_profile}b). (As noted in
\S\ref{sec:spix}, the region of the filament immediately adjacent to the
tail is affected by an image artifact and the spectral index there is not
reliable.)

%%%%%%%%%%%%%%%%%%%%%%%%%%%%%%%%%%%%%%%%%%%%%%%%%%%%%%%%%%%%%%%%%%%%%

%
%%%%%%%%%%%%%%%%%%%%%%%%%%%%%%%%%%%%%%%%%%%%%%%%%%%%%%%%%%%%%%%%%%%%%

\subsection{Radio spectrum of the BCG}\label{sec:bcgsp}

\begin{figure}
\centering
\includegraphics[width=8cm]{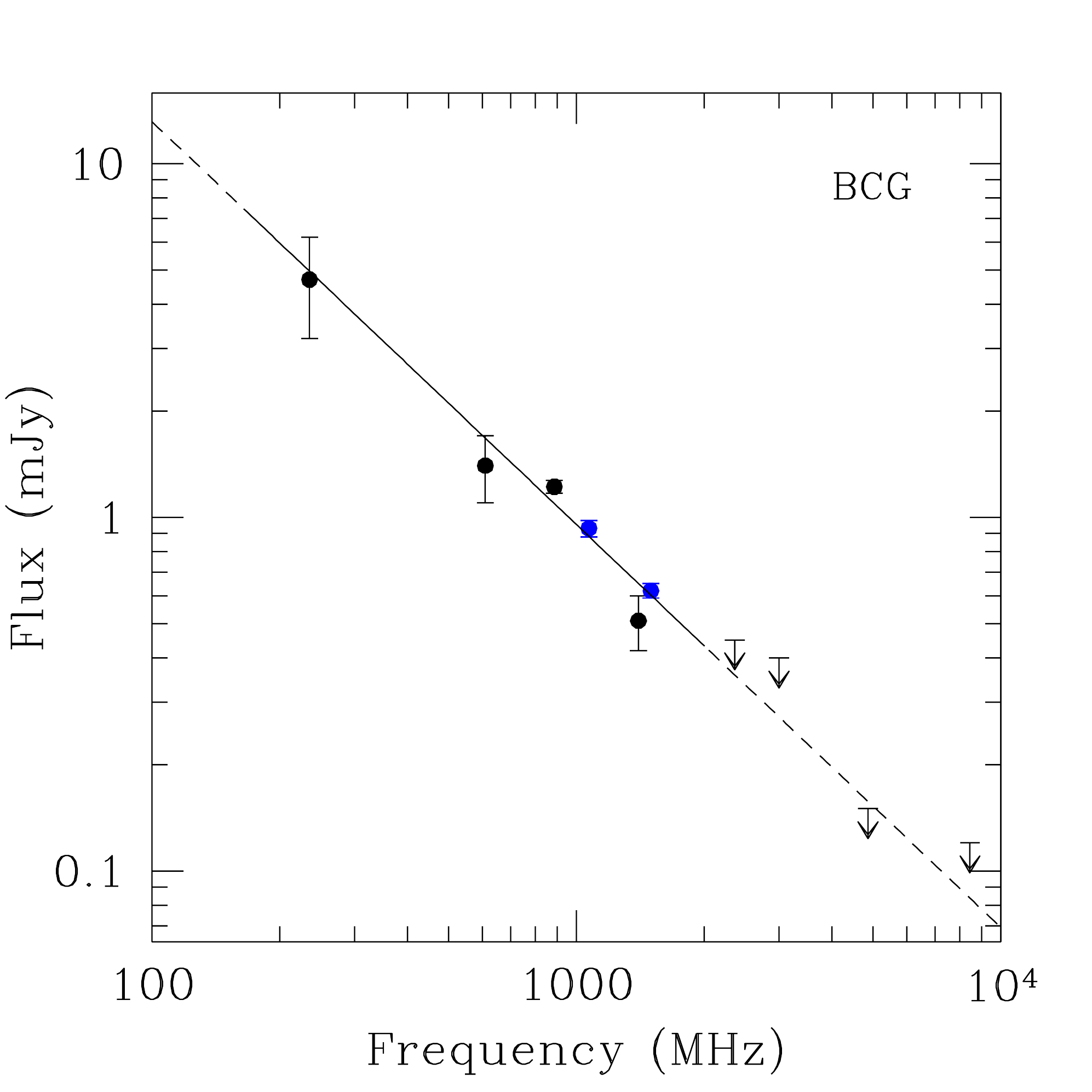}
\smallskip
\caption{BCG radio spectrum based on the flux densities listed in Table \ref{tab:bcg}. 
The MeerKAT flux densities, measured on the sub-band images at 1070 MHz and 1498 MHz, 
are shown in blue. The solid line is the best-fit power law model 
with $\alpha_{\rm tot}=-1.1\pm0.2$.} 
\label{fig:sp_bcg}
\end{figure}

Using the flux densities in Table \ref{tab:bcg}, we derived the 
radio spectrum of the BCG (Fig.~\ref{fig:sp_bcg}). There is some scatter, 
but overall the spectrum is steep with a single slope of $\alpha_{\rm tot}=-1.1\pm0.2$ over
the 235 MHz--1400 MHz interval (the solid line shows the best-fit power-law
model). Current $3\sigma$ upper limits at higher frequencies suggest a deviation from a
simple power-law behavior, suggesting a possible steepening above $\sim 2$ GHz.

We also measured the MeerKAT in-band spectral index for the whole BCG
emission and individual components. For the total emission, we obtained
$\alpha_{\rm tot}=-1.2\pm0.2$, consistent within the errors with the best-fit
slope in Fig.~\ref{fig:sp_bcg}. For the central compact component, we measured
$S_{\rm 1070, \, cc}=0.55\pm0.03$ mJy and $S_{\rm 1498, \,cc}=0.38\pm0.02$ mJy using 
a Gaussian fit, leaving $S_{\rm 1070, \, ext}=0.38\pm0.02$ mJy and 
$S_{\rm 1498, \, ext}=0.24\pm0.02$ mJy in extended emission. 
The resulting in--band spectral indices are both quite
steep, $\alpha_{\rm cc}=-1.10\pm0.23$ (central component) 
and $\alpha_{\rm ext} = -1.37\pm0.29$ (extended).

%
%%%%%%%%%%%%%%%%%%%%%%%%%%%%%%%%%%%%%%%%%%%%%%%%%%%%%%%%%%%%%%%%%%%%%

\section{X--ray data reduction}

X-ray {\em Chandra} and {\em XMM-Newton} images of A3562 have been presented
and analyzed in earlier works \citep[e.g., F04, G05,][V22]{2010A&A...516A..32G}.
In this paper, we 
re-analyze the {\em Chandra} observation to compare the newly-detected 
radio features with the brightness and temperature distributions of the 
thermal gas in the cluster center. We also use an {\em XMM-Newton} image 
and gas temperature map from V22.

\subsection{Chandra}

A3562 was observed by {\em Chandra} in 2003 (OBSID 4167) with ACIS-I for 20
ks. We reprocessed the Level-1 ACIS event file from the archive using CIAO
4.8, following the procedure described in \cite{2005ApJ...628..655V} using
the \chandra\ Calibration Database (CALDB) 4.8. Following
\cite{2003ApJ...586L..19M}, we excluded time intervals with elevated
background, which yielded a clean exposure of 18 ks. To model the detector
and sky background, we used the blank-sky datasets from the CALDB
appropriate for the date of the observation, normalized using the ratio of
the observed to blank-sky count rates in the 9.5--12 keV band. Following
\cite{2000ApJ...541..542M}, we also subtracted the ACIS readout artifact. We
obtained images in the 0.5--4 keV and 2--7 kev energy bands to detect unrelated
X-ray point sources, which were then masked during the image analysis.

\begin{figure*}
\centering
\includegraphics[width=8cm]{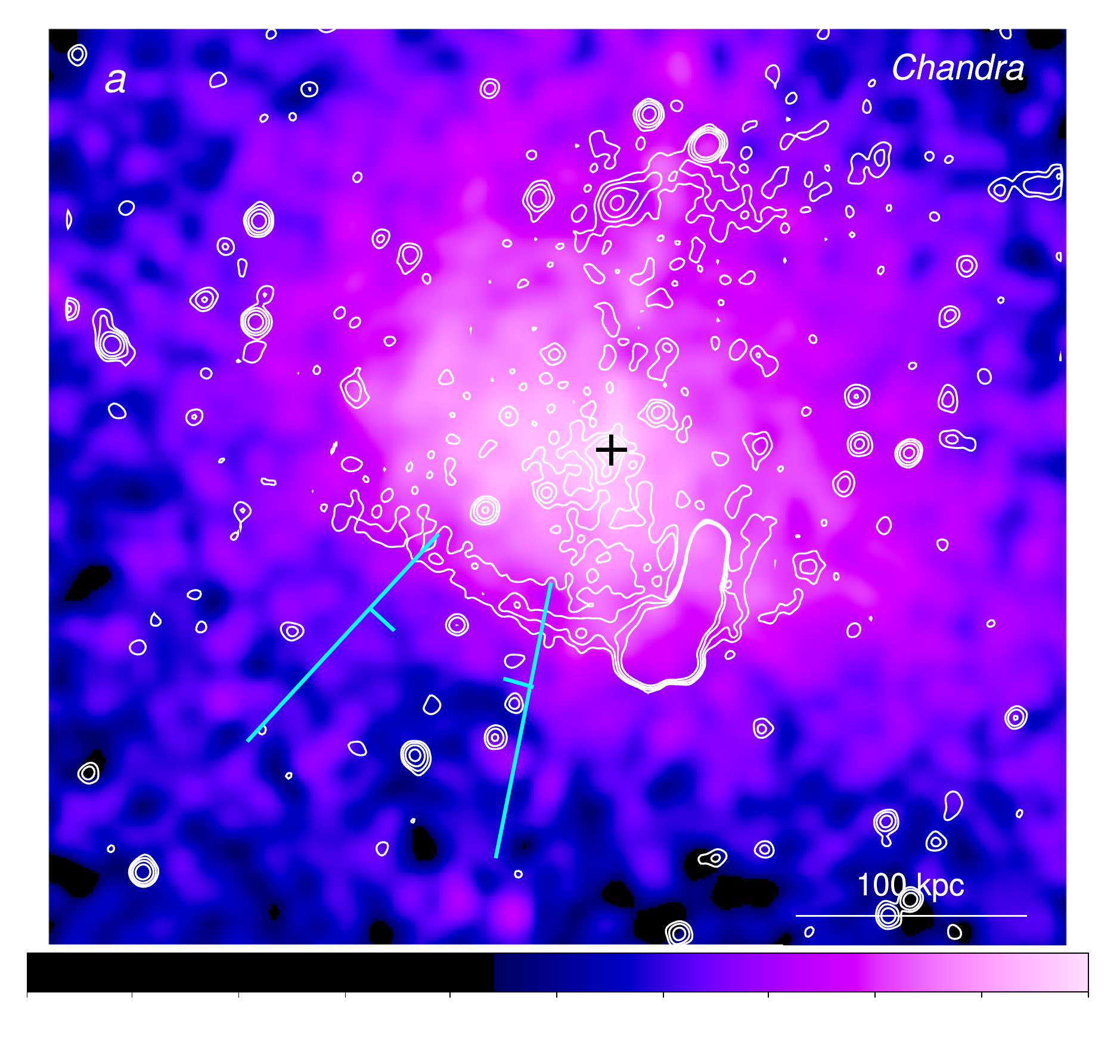}
\includegraphics[width=8cm]{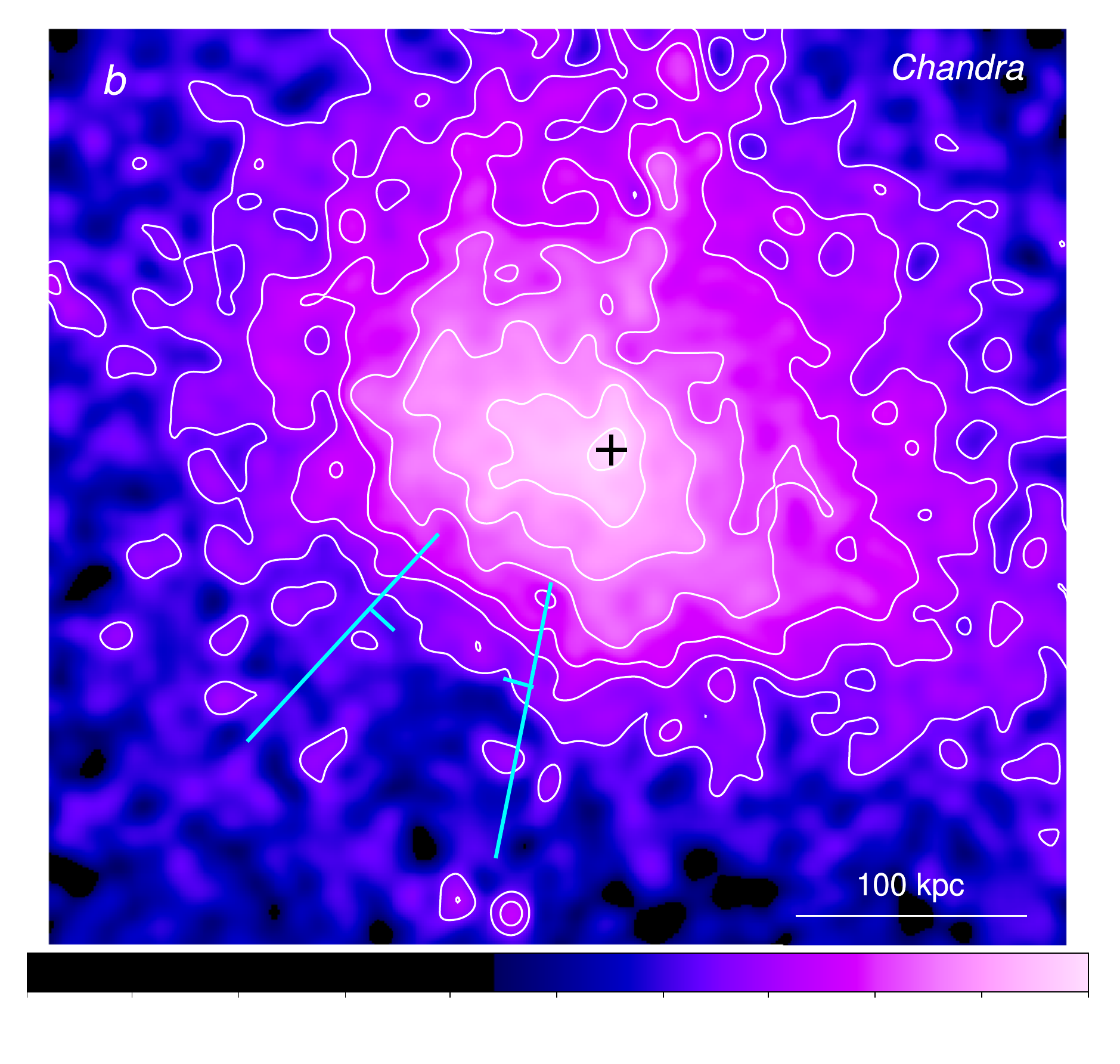}
\includegraphics[width=8cm]{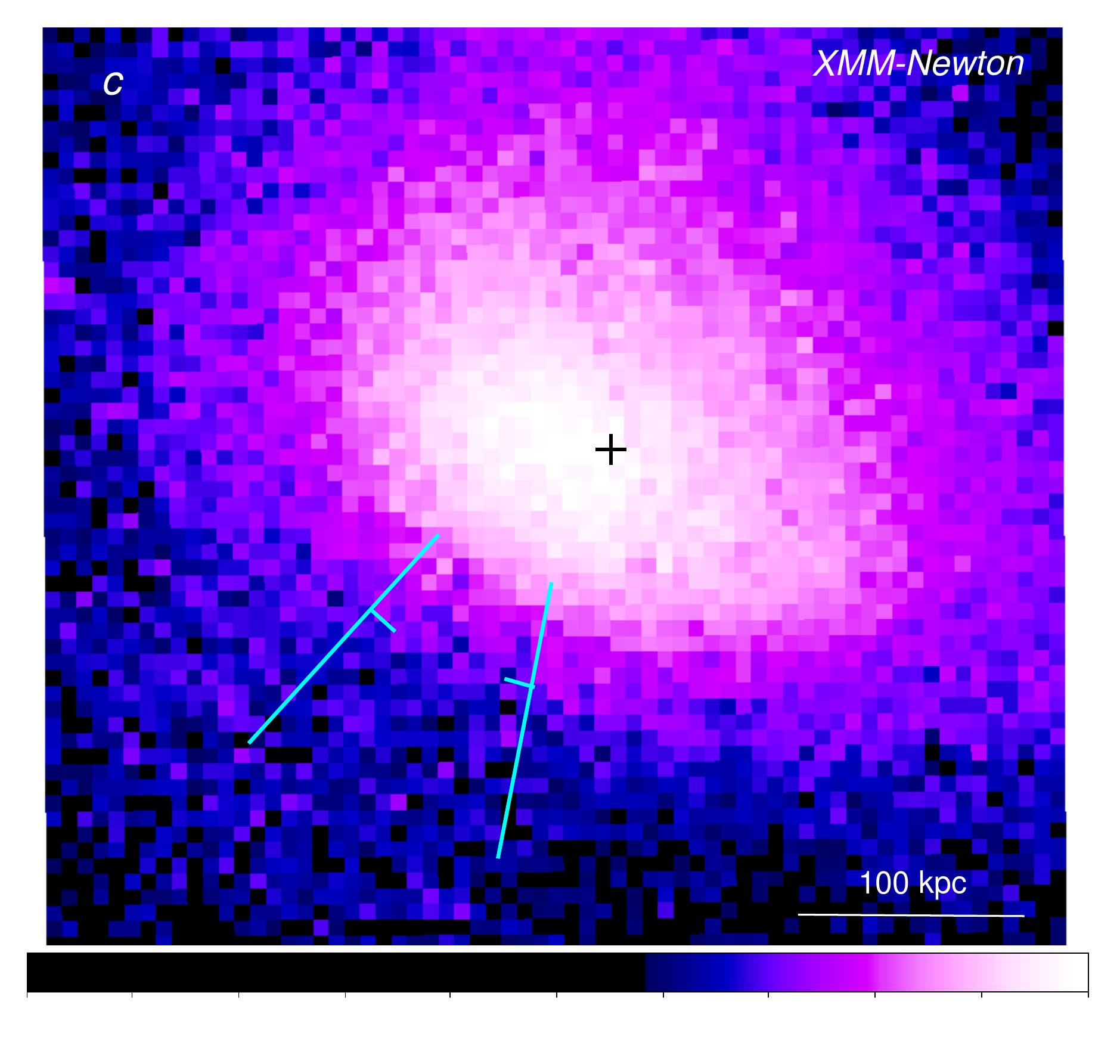}
\includegraphics[width=8cm]{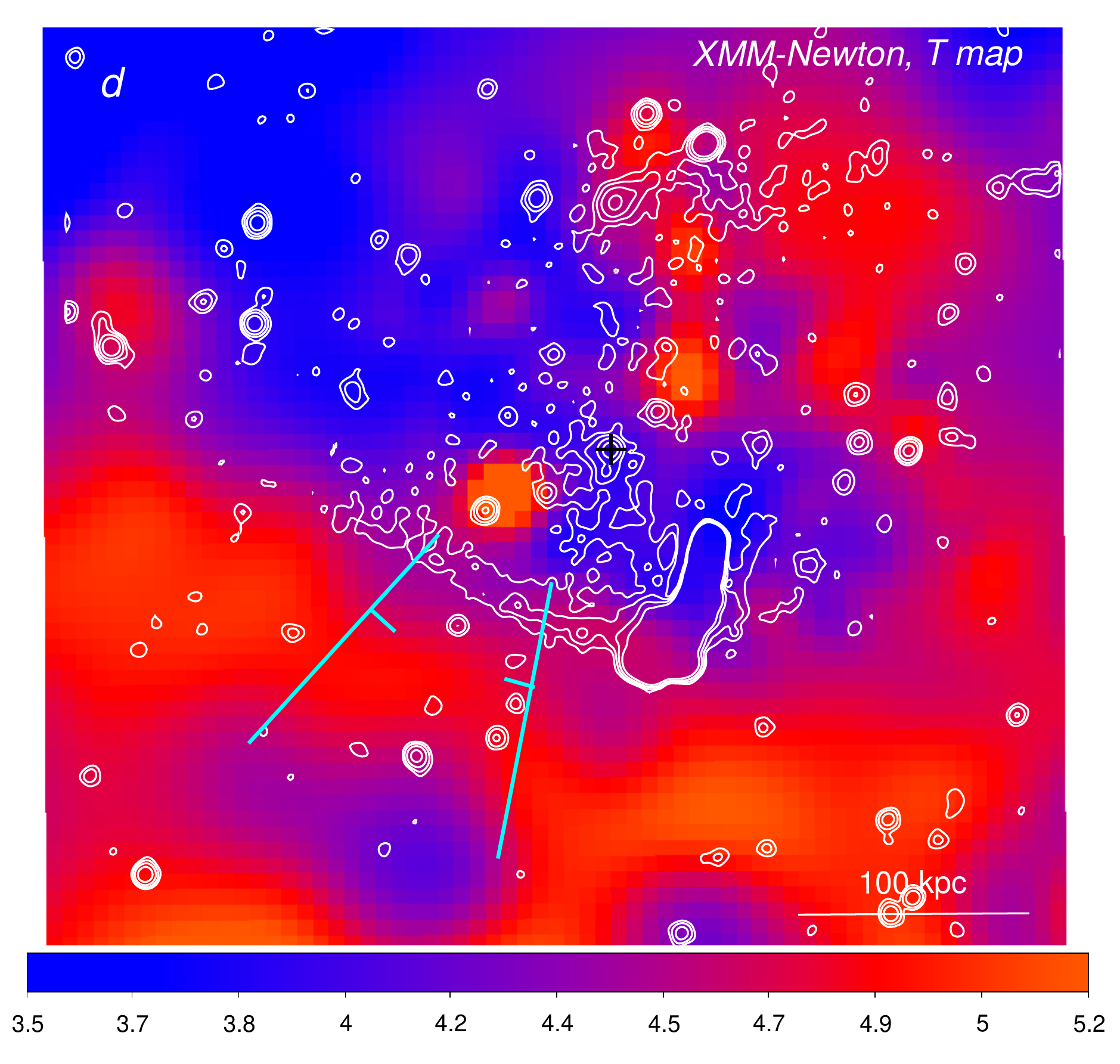}
\smallskip
\caption{({\em a}) {\em Chandra} image in the 0.5-4 keV band, smoothed with
a $5^{\prime\prime}$ Gaussian. The image is background subtracted and
divided by the exposure map. Radio contours from Fig.~\ref{fig:meerkat}a are
overlaid at 24 ($4\sigma$), 48, 96 and 192 $\mu$Jy beam$^{-1}$. ({\em b}) {\em
Chandra} 0.5-4 keV image as in ({\em a}) with contours of the core X-ray emission 
overlaid, spaced by a factor of $\sqrt{2}$. ({\em c}) {\em XMM-Newton} image
in the 0.5-2.5 keV band (V22). ({\em d}) {\em XMM-Newton} temperature map (V22). 
In all panels, the black cross marks the position of the BCG. The cyan sector is
used for the radial profiles (Fig.~\ref{fig:xrayprof}); short ticks show the
position of a cold front. Color bars show the X-ray surface brightness in 
arbitrary units ({\em a}, {\em b}, and {\em c}) and temperature in keV ({\em d}).}
  \label{fig:xray}
\end{figure*}

\begin{figure}
\centering
\includegraphics[width=10cm]{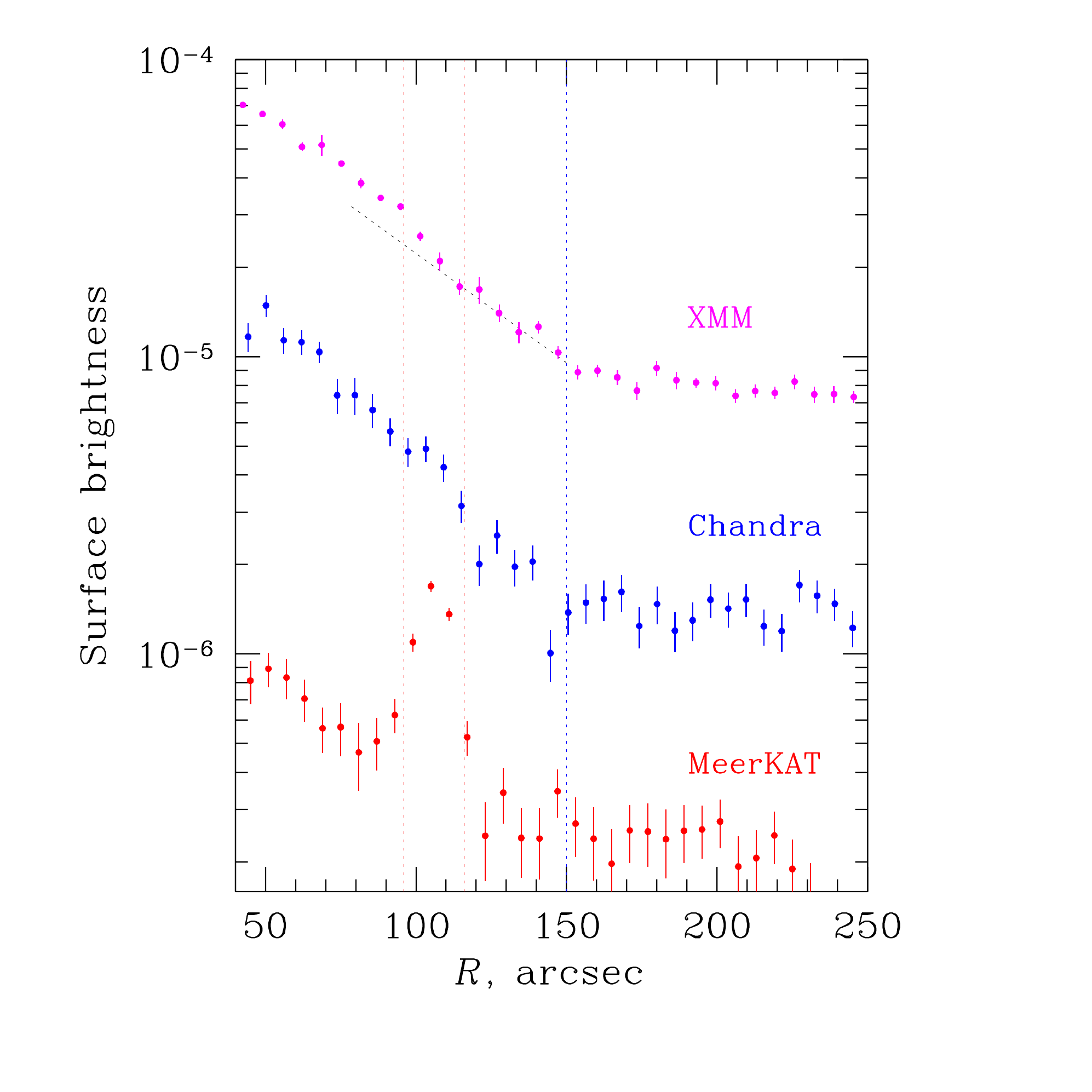}
\smallskip
\caption{Surface brightness profiles (arbitrary units) for MeerKAT (radio,
red), {\em Chandra} (X-ray, blue) and {\em XMM-Newton} (X-ray, magenta)
images, extracted in the sector shown in Fig.~\ref{fig:xray}. The radius is
from the cluster X-ray centroid. Blue dotted line shows the position of the
X-ray cold front \citep{2010A&A...516A..32G}.  Red dotted lines show the radio
filament. A hint of a corresponding X-ray front feature is suggested inside
the main cold front, though its significance is low. The {\em XMM} profile
suggests a break at that position consistent with an additional edge 
(see black dotted line that matches the profile between the X-ray cold 
front and the outer edge of the filament).}
  \label{fig:xrayprof}
\end{figure}

\section{X-ray and radio analysis}

In the X-ray band, A3562 is known to exhibit a very disturbed atmosphere
\citep[e.g.,][F04]{2002A&A...382...17B}. It does not possess a central cool core,
but rather a {\em warm} core \citep[F04][]{2009ApJS..182...12C}, with a 
specific entropy lower than in the surrounding gas.  The core appears to be
sloshing in response to the passage of the galaxy group SC 1329–313
(currently southwest of A3562) to the North of the cluster (F04), and a cold
front is seen South-West of the center \citep{2010A&A...516A..32G}.

In Fig.~\ref{fig:xray}{\em a}, we overlay the MeerKAT high-resolution image
contours on the 0.4-5 keV {\em Chandra} image of the cluster central
region. Panels {\em b} and {\em c} show the same {\em Chandra} image and an
{\em XMM-Newton} image in the 0.5-2.5 keV band (V22).  A map of the projected gas
temperature from the {\em XMM} data is shown in Fig.~\ref{fig:xray}{\em d}
(V22).  The core X-ray emission is highly asymmetric with respect to
the BCG location (black cross) and is more extended toward North,
as previously noted by F04. The radio
filament is oriented tangentially to the cluster center, and follows 
the isophotes of the X-ray core.

We used the cyan sector marked in the images in Fig.~\ref{fig:xray},
which encompasses the best-defined part of the radio filament, to
extract radial profiles of the radio and X-ray surface brightness. 
The sector is centered on the centroid of the large-scale cluster X-ray 
emission, which is offset from the X-ray peak. It also approximately coincides 
with the center of curvature of the X-ray edge and thus our profiles describe 
it accurately.
The profiles are shown in Fig.~\ref{fig:xrayprof}. The radio profile (red) was
extracted on the high-resolution image in Fig.~\ref{fig:meerkat}a and shows
a clear brightness enhancement at the position of the radio filament (red dotted
lines). A brightness edge is apparent in the X-ray profiles (marked by the blue
dotted line), particularly with {\em Chandra}'s angular resolution; its
position is also marked with cyan ticks in the images in
Fig.~\ref{fig:xray}. The temperature map in Fig.~\ref{fig:xray}{\em d}, as
well as a temperature map in F04, show that the gas inside this edge is
cool, so the X-ray edge is a cold front. A similar conclusion was reached
based on the radial surface brightness and temperature profiles by \cite{2010A&A...516A..32G}\footnote{
We note that \cite{2010A&A...516A..32G} used a different center 
to extract the X-ray profile (their Fig.~A.10). The X-ray edge 
at $100^{\prime\prime}$ from their center (13h33m36.8s,$-$31d40$^{\prime}$20.4$^{\prime\prime}$)
coincides with the position of the edge in Fig.~\ref{fig:xrayprof} (blue dotted line), 
at $150^{\prime\prime}$ from our center (13h33m33.9s,$-$31d39$^{\prime}$51.1$^{\prime\prime}$).}.
From the X-ray images, the X-ray edge is much wider than the sector in
Fig.~\ref{fig:xray}, which we selected to highlight the radio filament. Cold
fronts are often seen in cluster sloshing cores, and sloshing is easily set
off by any cluster merger as long as there is a negative radial entropy
gradient \citep{2007PhR...443....1M}, such as we have in A3562 (F04).

The radio filament is parallel to the cold front. However, it is not located
at the front but $20^{\prime\prime}-30^{\prime\prime}$ inward, as seen from
the radio brightness profile in Fig.~\ref{fig:xrayprof}. The {\em Chandra}\/
profile hints at another edge-like X-ray feature (or a bump) at the position
of the radio filament. The {\em XMM}\/ point spread function dilutes any
sharp features (as indeed seen for the main cold front), but the {\em XMM}\/
profile also suggests an inflection at that position, consistent with the
presence of another edge (see black dotted line in Fig.~\ref{fig:xrayprof}).
If this feature is real, it could be a wiggle on the surface of the cold
front seen in projection, or a secondary density edge inside the main
front. It may also be a combination of gas {\em depletion}\/ layers at the
location of strong magnetic field filaments \citep{2007PhR...443....1M, 2011ApJ...743...16Z,2016MNRAS.455..846W}.
The possibility that this X-ray bump is inverse Compton emission 
from the radio filament does not appear to be plausible as the much 
brighter source shows no hint of such X-ray excess. 
A forthcoming deeper {\em Chandra}\/ observation will clarify
whether this X-ray feature is indeed associated with the radio filament.

\section{Discussion}

\subsection{Origin of the radio filament}\label{sec:disc1}

The new MeerKAT image has revealed a narrow and faint radio filament
departing from the end of the J1331-3141 head tail and stretching
approximately 200 kpc eastward. While we observe a clear spectral steepening
along the head tail, indicating aging of the relativistic electrons as
they move away from the injection site, the spectral index along the
filament is instead relatively uniform around $\alpha ~\sim -1.5$
(although with large uncertainties), which is the value seen in the tail at
the position where the filament branches out. This uniformity of the
spectrum is similar to that of the MeerKAT radio threads between the lobes of 
ESO 137--006 \citep{2020A&A...636L...1R} as well as the LOFAR structures in 
A1033 \citep{2017SciA....3E1634D}, where they also appear to originate 
from a head-tail source.

\begin{figure}
\centering
\noindent
\vspace*{3mm}

\includegraphics[width=7.0cm]{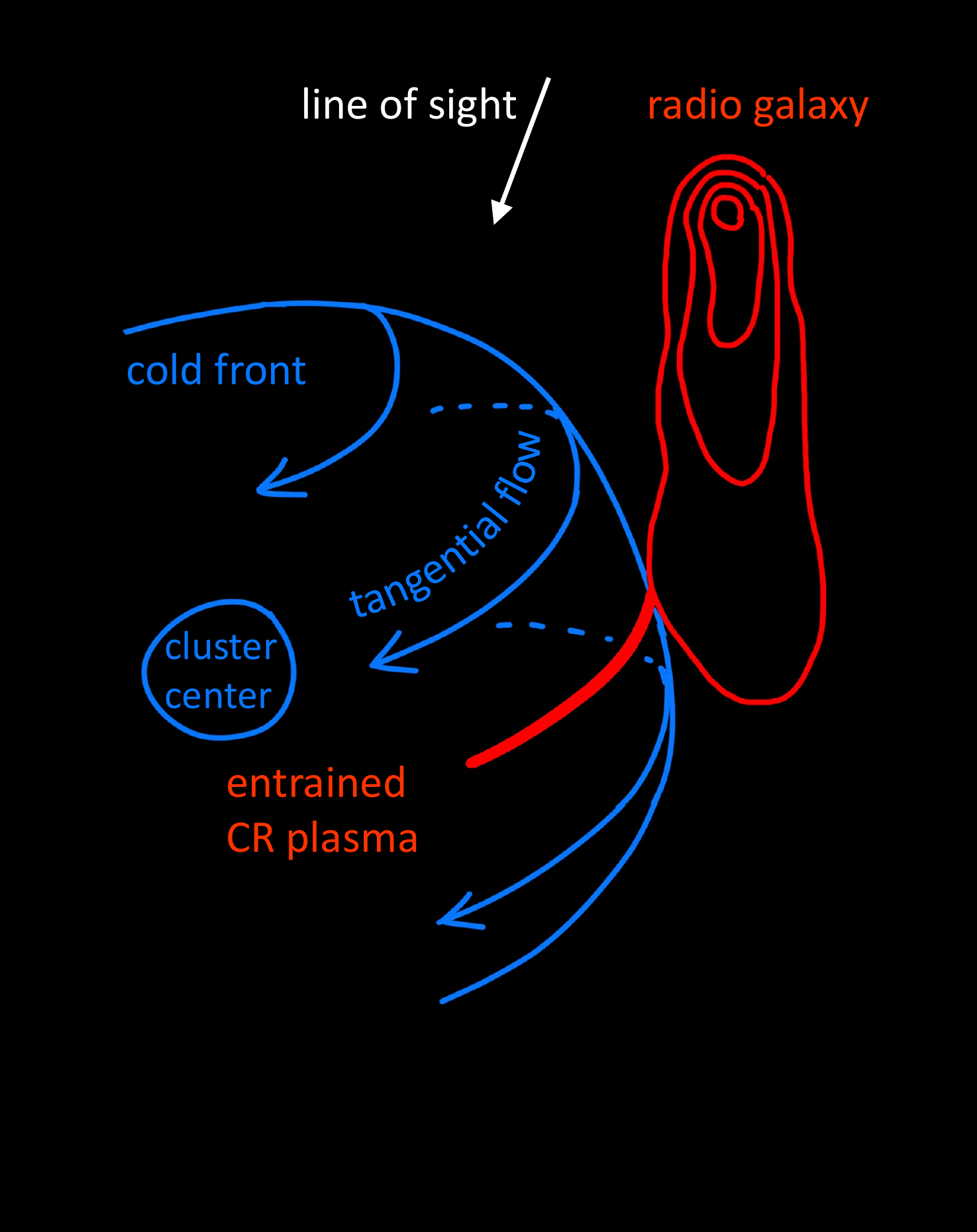}

\vspace*{-4mm}\hspace*{0mm}
\includegraphics[width=7.0cm]{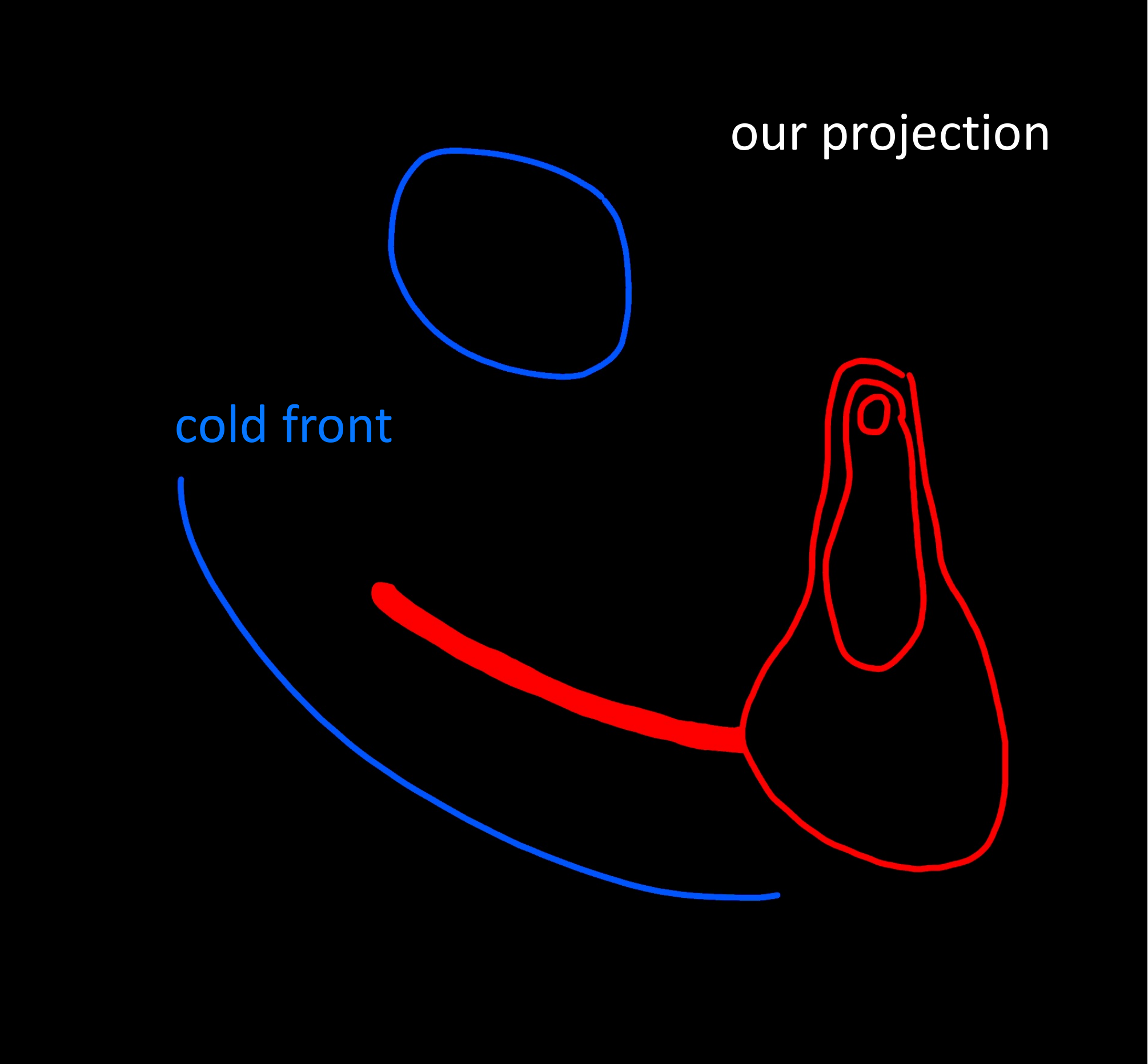}
\smallskip

\caption{A possible geometry where of the radio galaxy is located outside
the cold front but dips its tail under the front. The tangential wind under
the front entrains a parcel of the cosmic ray-enriched gas from the tail and
stretches it along the flow lines (alternatively, the cosmic ray electrons diffuse
along the magnetic field lines stretched by the gas flow). The lower panel
shows our image plane.}
  \label{fig:sketch}
\end{figure}

We found that the radio filament is parallel to an X-ray cold front seen
$\sim 100$ kpc southwest of the cluster center, but located $\sim 20-30$ kpc
behind it (towards the cluster) in projection. Numerical simulations of gas
sloshing in cluster cores predict prevalent tangential flow of gas under the
sloshing cold fronts \citep{2011ApJ...743...16Z}. If a parcel of gas, enriched
with cosmic ray (CR) particles from the head-tail source, gets under the front,
the {\em wind}\/ there would pick it up and stretch it along the
front. The tail region of the radio galaxy is advected by the 
motions of the surrounding cluster plasma. If it were located in the sloshing 
core, it would be quickly disrupted by the large bulk flows there 
(e.g., ZuHone et al. 2021). Because it does not appear to feel the wind, 
the head-tail source is probably outside the dense core and the front.
In 3D, cold fronts are convex structures roughly concentric with
the cluster X-ray peak.  We suggest that the head-tail source is located
just outside the front, but projected onto the inside region, with its tail
touching the front surface and dipping into a tangential stream of gas under
the front, as shown in Fig.\ \ref{fig:sketch}. While the region inside the
cold front should be isolated thermally and magnetically from the gas
outside the front by draping of the field, it is not perfect and some field
lines cross the front \citep{2013ApJ...762...69Z}. The base of the radio 
filament may be at such a location, where the cosmic ray electrons from 
the tail may have diffused under the cold front along such stray lines.

%%%%%%%%%%%%%%%%%%%%%
%%%%%%%%%%%%%%%%%%%%%%

In this picture, the uniformity of the spectral index along the radio filament 
is explained naturally, because all locations along the filament have the same
relative radiative age --- the parcel of gas of the same initial age is
stretched fast by the flow, as opposed to fresh electrons being fed to one
end of the tube and slowly traveling along the tube as they age, as in the
head-tail source.  A problem with this picture is the length of the filament
($\sim 100$ kpc for the interval where the spectral index can be measured)
and the speed of the wind ($\sim 500$ km\,s$^{-1}$ for a Mach $M=0.5$ flow),
which give the timescale for stretching of $\sim 200$ Myr. During such time,
the electrons would (uniformly) age out of the MeerKAT radio band. Thus,
reacceleration of the electrons in the filament may be needed.
Reacceleration of aged electrons (originating from the head tail) by
turbulence has been suggested for the A1033 filaments \citep{2017SciA....3E1634D}
that have been named GReET (gently re-energized tail).  A very
gentle re-energization process is in fact required to explain the brightness
and spectral properties of these filaments with $\alpha \approx -4$, in
which the particle reacceleration timescale is comparable to the radiative
loss timescale of the electrons emitting at LOFAR frequencies. Other
cluster radio tails, showing a low-frequency brightness and spectral 
behavior that is inconsistent with aging models, may be further examples 
of this gentle re-energization mechanism at work \citep{2018A&A...609A..61C,
2018MNRAS.473.3536W, 2021A&A...649A..37B, 2021MNRAS.508.5326M, 2021arXiv211115517I}. 
However, this process may not be sufficient to sustain the radio emission seen 
at GHz frequencies from the A3562 filament, whose radio spectrum 
($\alpha\sim -1.5$) is much flatter than, for instance, in A1033.

Alternatively, we may have a much faster stretching. The same tangential
flow inside the cold front also stretches and amplifies the normally tangled
magnetic field into filaments parallel to the cold front \citep{2011ApJ...743...16Z}.
If the relativistic electrons from the same patch of the tail
that dipped into the flow could diffuse along the field lines (rather than
simply being advected by the flow) much faster than the Alfven speed
(typically $\lax 100$ km\,s$^{-1}$, probably higher in the amplified field
under the cold front) and faster than the gas flow ($\sim 500$
km\,s$^{-1}$), it may solve the aging problem. However, there is still 
theoretical debate about how fast the cosmic rays can diffuse in the ICM 
\citep[e.g.,][]{2011A&A...527A..99E,2013MNRAS.434.2209W,2018MNRAS.473.3095W}.

The filament delineates the southern edge of the brighter region of the radio halo that fills the cluster core, and it is interesting to compare it with other examples of such a spatial coincidence. There have been observations of long radio relics located at the edges of giant radio halos, or sharp edges in the radio-halo emission, some of which coincide with a shock front seen in the X-ray (e.g., the relic at the edge of the halo in A521, \cite{2008A&A...486..347G}, \cite{2008Natur.455..944B}, \cite{2013ApJ...764...82B} and in A\,754, \cite{2011ApJ...728...82M}; the radio-halo edges in the Bullet cluster, A\,520 and Coma, \cite{2005ApJ...627..733M}, \cite{2014A&A...561A..52V}, \cite{2014MNRAS.440.2901S}, \cite{2018ApJ...856..162W}, \cite{2019A&A...622A..20H}, \cite{2011MNRAS.412....2B}, \cite{2022arXiv220301958B}; see \cite{2019SSRv..215...16V} for other examples). The physical interpretation there is the shock passage re-energizing the fossil relativistic electrons and igniting a relic, and the same gas inflow that produces the shock front also producing turbulence downstream from the shock (toward the cluster center), which in turn re-accelerates the cosmic ray electrons which emit the radio halo. 
In principle, it is possible that the filament in A3562 is a relic on top of a shock at the edge of the halo. However, in the X-ray, we observe a cold front near and parallel to the filament, rather than a shock front. So such a scenario is unlikely, unless the subtle second X-ray brightness edge suggested at the location of the filament (Fig.\ 9) is in fact a shock front outside the core seen in projection onto the core. That edge and the radio filament appear to be very closely concentric with the main cold front (marked by ticks in Fig. 8), which would have to be a complete coincidence, and the 3D geometry of the shock would have to be very unlikely (merger shock fronts in clusters tend to propagate in the radial direction, while this one would have to propagate tangentially). Our forthcoming longer X-ray observation with {\em Chandra} will clarify the nature (and confirm the existence) of that second X-ray edge. Given the existing data, our interpretation for the filament to arise from stretching by the gas flow under the cold front is more plausible.

Future spectral and polarization radio studies will help to understand the
nature of the filament, and potentially provide constrain 
on the physics of the ICM. In particular, sensitive and high-resolution 
radio observations of A3562 at lower frequencies (e.g., with the uGMRT) are
needed to reduce the uncertainties on the spectral index along the 
radio filament. Polarization observations will be crucial to estimate 
the magnetic field strength and structure. Unique to the relics at the shock fronts, there is a steepening of the radio spectrum observed across the relic, from the front downstream (e.g., \cite{2008A&A...486..347G}, \cite{2019SSRv..215...16V}). If such transverse steepening is observed in the A3562 filament, this would strongly argue for a shock scenario rather than our preferred explanation. This test would require high-sensitivity observations with a higher angular resolution than currently available.

\subsection{Origin of radio halo}

As discussed above, the radio filament may not mark the location of a shock front, but it may still connected to the radio halo. V03 suggested that the head-tail
radio source may supply the cosmic ray electrons that feed the radio halo,
where they are then reaccelerated, e.g., by turbulence, to counterbalance
the synchrotron and inverse-Compton energy losses. The newly-discovered radio
filament may be showing us how the seed CR electrons are channeled from the
head-tail into the cluster core. The magnetic field structure inside the
cold front is interconnected (and largely isolated from that outside the
front, \cite{2013ApJ...762...69Z}). The electrons from the filament can diffuse
throughout the gas inside the front along those connected lines. 
The radio tail may have provided more seed electrons to the cluster core 
than those channeled through this filament. It is
possible that there are (or were) other such magnetic crossings of the cold
front between the tail and the inner gas, perhaps not always resulting in
such highly visible filaments. The turbulence needed to reaccelerate 
electrons within the core can be produced by sloshing in
response to SC group passage (G05), as suggested for radio 
minihalos by \cite{2013ApJ...762...78Z}. In the latter work, 
the seed electrons were proposed 
to come from disrupted radio lobes of the central radio galaxy; 
those seeds could be supplied by an external radio galaxy 
as well \citep[see also][]{2021ApJ...911...56G,2021MNRAS.508.3995B}.

\subsection{Southern radio extension}

We find very faint diffuse radio emission outside the brighter part of the
radio halo, in particular, southwest of core, outside of the radio filament and
the cold front (``South emission'' in Fig.\ \ref{fig:meerkat}{\em b}). It
appears related to the X-ray cluster structure as well --- F04 noticed an
azimuthally asymmetric enhancement of the X-ray brightness and
pseudo-pressure in that region southeast of center just outside the cold
front (see their Figures 3 and 5). A similar association (at least in projection) with a higher
pseudo-pressure region was reported by G05 for the Western extension
(Fig.\ \ref{fig:meerkat}{\em b}), which we now see merging at larger
distances into the Mpc-scale bridge connecting the halo and the 
SC 1329--313 group (V22). The cluster appears to
be full of diffuse radio emission at all linear scales, with the brighter
region inside the sloshing core with an abrupt boundary at the radio filament ---
possibly another example of a multiple-component radio halo with a bright
minihalo, or minihalo-like emission, in the sloshing core that co-exists 
with a fainter diffuse component outside the core
\citep[e.g.,][]{2015MNRAS.448.2495S,2017A&A...603A.125V,2018MNRAS.478.2234S,2021MNRAS.508.3995B,2021arXiv211207364R}.

\begin{figure}
\centering
\includegraphics[width=8cm]{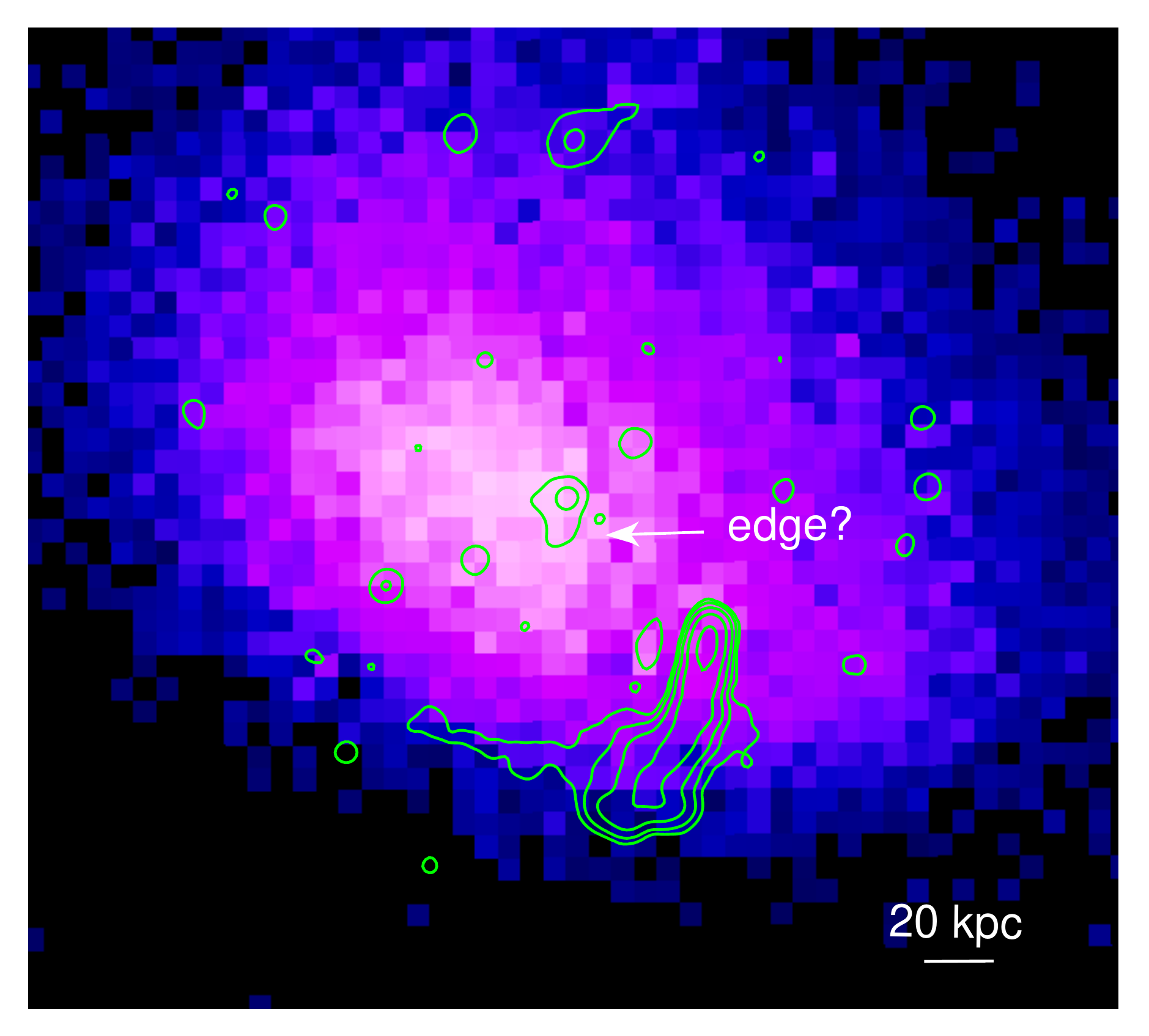}
\smallskip
\caption{The core of A3562. The same {\em XMM}\/ X-ray image as in Fig.\ 8c,
with the colors chosen to emphasize the possible X-ray brightness edge near
the central galaxy. Radio contours are as in Fig.\ 3.}
  \label{fig:bcgxmm}
\end{figure}

\subsection{A radio-quiet BCG}

The gigantic BCG of A3562 (Fig.\ 3) hosts a very faint radio source with a
radio luminosity of $4\times10^{21}$ W Hz$^{-1}$ (Sect.~\ref{sec:bcg}) only,
namely $2-3$ orders of magnitude lower than cD galaxies with AGN in
clusters of a similar mass \citep[e.g.,][]{2015MNRAS.453.1201H}. 
The cD galaxy in the massive, cool-core cluster Ophiuchus has a similar radio-faint nucleus
\citep{2010A&A...514A..76M,2016MNRAS.460.2752W,2020ApJ...891....1G}. 
For Ophiuchus, the possible reason was revealed by a high-resolution {\em
Chandra}\/ image, which shows small-scale sloshing cold fronts and a 2 kpc
offset of the peak of the dense gas from the nucleus \citep{2016MNRAS.460.2752W};
the peak of the molecular gas shows a similar offset \citep{2012MNRAS.421.3409H}. 
A possible interpretation of these offsets is that the ongoing gas sloshing
has temporary displaced the densest and coolest gas, interrupting the supply
of the accretion fuel for the nucleus and thus weakening its radio activity.

While A3562 does not possess a cool core, it has low-entropy gas in its
center (F04) that could provide fuel to the radio source, had it been focused 
on the BCG nucleus. The BCG radio spectrum is quite 
steep ($\alpha=-1.1$, \S\ref{sec:bcgsp}), suggesting the radio source is 
old and possibly unfed.  The {\em XMM}\/ image in Fig.~\ref{fig:bcgxmm} shows that the gas
density peak may in fact be offset from the BCG, likely by the same sloshing
motions that have created the cold front southwest of the cluster
center. The {\em XMM}\/ image may even hint at the presence of another cold
front near the BCG (see also Fig.~\ref{fig:xray}{\em c}, where a vertical
edge is seen at the position of the cross), as in Ophiuchus, though the
image resolution is not sufficient and there is also a chip artifact.  The
forthcoming {\em Chandra}\/ observation should provide a better picture of
the gas around the BCG. This stage in the BCG activity cycle is 
expected to be transient and extremely rare \citep{2012MNRAS.421.3409H}, 
as only few cluster AGNs with clear offset gas peaks are known to date.

\section{Summary}

A MeerKAT observation of the A3562 galaxy cluster revealed a curious narrow,
straight synchrotron filament that branches out at a straight angle from the
tail of a radio galaxy and spans almost 200 kpc. Its surface brightness is
much lower than that of the tail. The radio filament bounds the bright region of
the previously known radio halo that fills the cluster. Comparing the radio
image with the \chandra\ and \xmm\ X-ray images, we discover that the
filament traces a cold front that delineates the sloshing cluster core ---
but it stays inside the front in projection.

While the radio galaxy tail shows clear spectral steepening from the head
along the tail, the filament shows a rather constant spectral index of
$\alpha\simeq -1.5$ (although with large uncertainties). This is also the spectral 
index of
the tail at the distance where the filament is emerging. We speculate
that the filament forms where the tail of the radio galaxy, located outside
of the sloshing core and the cold front, touches the front. The fast
tangential gas flow under the front can stretch the parcel of plasma enriched
with cosmic ray electrons into a narrow filament. It would be natural to
have a uniform spectral index along the filament because the age of the
electrons is the same along the filament. Some reacceleration is needed 
in this scenario to maintain the radio filament detectable at the MeerKAT 
frequencies. Alternatively, we may see
anomalously fast diffusion of those electrons from the tail along the
magnetic field lines stretched by the gas flow. Our newly discovered
synchrotron filament offers an interesting experimental setup to constrain this process.

The radio filament may have been channeling seed electrons from the tail into the
region under the cold front, where they are reaccelerated by turbulence and
form the radio halo. We also detect radio emission beyond (outside) the
filament and the cold front, but its surface brightness is abruptly lower
than that inside the filament-bound halo region.

Finally, we detect a faint radio source at the center of the BCG. It is 2--3
orders of magnitude fainter than the typical central BCG in clusters of
similar mass. It is likely that the radio source is currently starved 
of accretion fuel by core sloshing, as seen in a few other clusters such 
as Ophiuchus.

\vspace{5mm}
{\it Acknowledgements.}
We thank the referee for their critical and helpful comments. Basic research in radio astronomy at the 
Naval Research Laboratory is supported by 6.1 Base funding. The MeerKAT telescope is operated by the South 
African Radio Astronomy Observatory, 
which is a facility of the National Research Foundation, an agency of the Department of Science and Innovation. The authors acknowledge the contribution of those who designed and built the MeerKAT instrument. The financial assistance of the South African Radio Astronomy Observatory (SARAO) towards this research is hereby acknowledged (www.sarao.ac.za).
The Australian SKA Pathfinder is part of the Australia Telescope National Facility which is managed by CSIRO. Operation of ASKAP is funded by the Australian Government with support from the National Collaborative Research Infrastructure Strategy. ASKAP uses the resources of the Pawsey Supercomputing Centre. Establishment of ASKAP, 
the Murchison Radio-astronomy Observatory and the Pawsey Supercomputing Centre are initiatives of the Australian Government, with support from the Government of Western Australia and
the Science and Industry Endowment Fund. We acknowledge the Wajarri Yamatji people as the traditional owners of the Observatory site. 
The National Radio Astronomy Observatory is a facility of the National Science Foundation operated
under cooperative agreement by Associated Universities, Inc. The GMRT is run by the National Centre for Radio Astrophysics of the Tata Institute of Fundamental Research.
 T. Venturi and G. Bernardi acknowledge the support from the Ministero degli Affari Esteri e della Cooperazione Internazionale, Direzione Generale per la Promozione del Sistema Paese, Progetto di Grande Rilevanza ZA18GR02. H. Bourdin and P. Mazzotta acknowledge financial contribution from the 
contracts ASI-INAF Athena 2019- 27-HH.0, “Attivit{\`a} di Studio per la comunit{\`a} scientifica di Astrofisica delle Alte Energie e Fisica Astroparticellare” (Accordo Attuativo ASI-INAF n. 2017-14- H.0), from the European Union’s Horizon 2020 Programme under the AHEAD2020 project (grant agreement n. 871158) and support from INFN through the InDark initiative. The scientific results reported in this article are based data obtained from the Chandra Data Archive (ObsID  4167). 
This research has made use of software provided by the Chandra X-ray Center (CXC) in the application package CIAO.

{}

\appendix

\section{Inter-cluster diffuse radio emission in the A3562/SC1319-313 region}

\begin{figure*}
\centering
\includegraphics[width=\hsize]{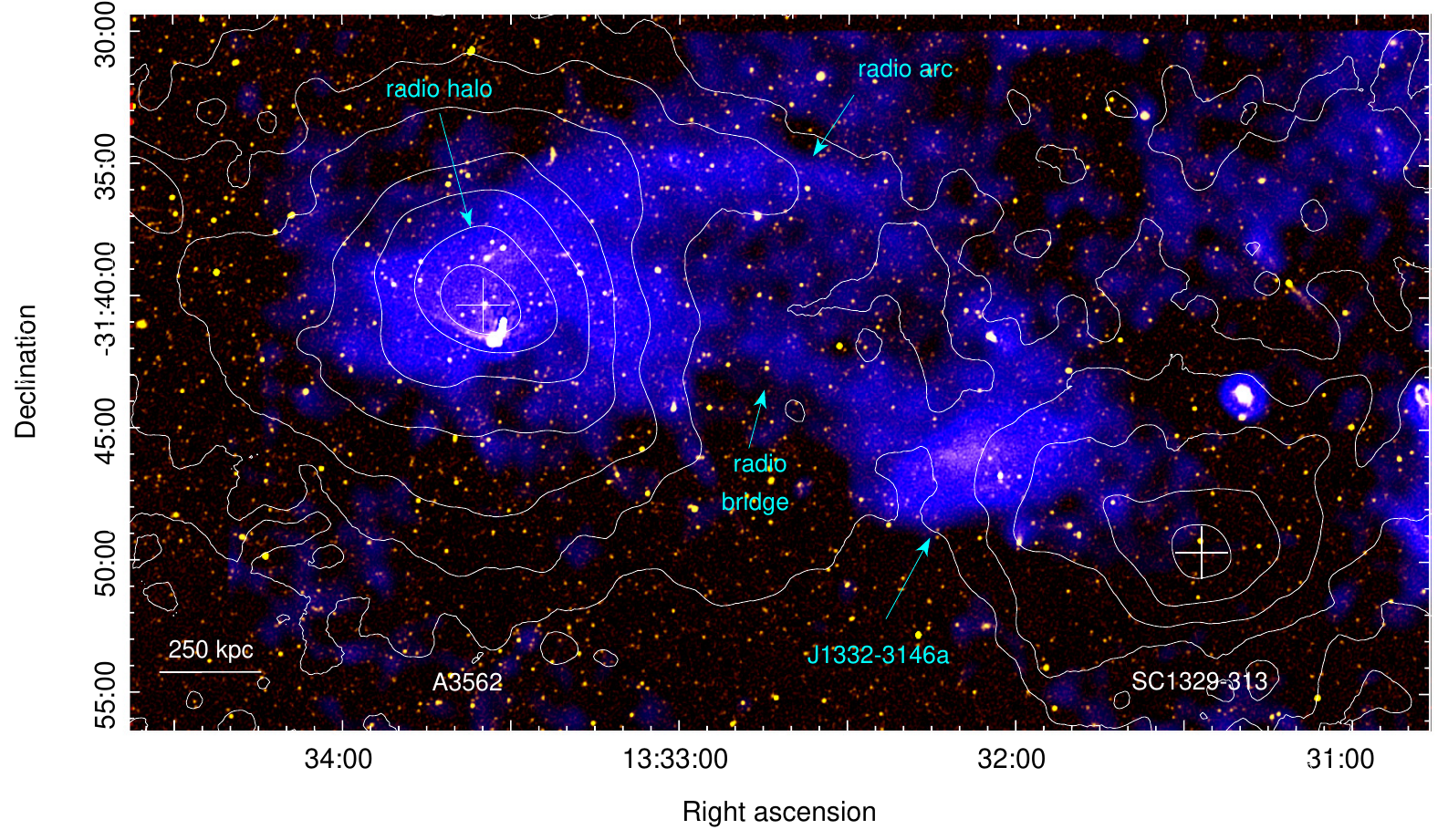}
\smallskip
\caption{Composite radio/X-ray image of the region of A3562 and nearby galaxy group SC1329--313. 
Colors show the MeerKAT image at $6^{\prime\prime}$ resolution (yellow; same as in
Fig.~\ref{fig:meerkat}{\em a}) and the MeerKAT image of the diffuse emission 
at $40^{\prime\prime}$ resolution (blue; from V22). 
The {\em XMM-Newton} X-ray image in the 0.5-2.5 keV band (V22) is shown as contours, spaced by a factor of 2.
White crosses mark the X-ray centers of A3562 and SC1329--313.}
\label{fig:arc}
\end{figure*}

In this appendix, we show a composite MeerKAT/{\em XMM-Newton} image of a 
2.7 Mpc $\times$ 1.5 Mpc region encompassing the galaxy cluster A3562
and nearby galaxy group SC1329--313 (Fig.~\ref{fig:arc}). The X-ray emission
is shown as contours. The MeerKAT 1284 MHz image at high resolution 
(Fig.~\ref{fig:meerkat}{\em a}) is shown in yellow color. In blue, 
we report the MeerKAT 40$^{\prime\prime}$-resolution image of the large-scale 
diffuse emission obtained by V22 after filtering out the contribution of individual 
sources. Two Mpc-scale, faint radio structures, labelled radio arc and 
radio bridge, are visible in the region between A3562 and SC1329-313. 
These structures appear to connect the radio halo in A3562 and diffuse radio source 
J1332–3146a at the outskirts of the SC group (V22).

%
%%%%%%%%%%%%%%%%%%%%%%%%%%%%%%%%%%%%%%%%%%%%%%%%%%%%%%%%%%%%%%%%%%%%%

\end{document}